\documentclass[english,preprint,aps,prb,preprintnumbers,floatfix]{revtex4}

\makeatletter

\usepackage{graphicx}
\usepackage{epstopdf}
\usepackage{babel}
\usepackage{epsfig}

\makeatother

\begin{document}

\title{Role of electrostatic forces in cluster formation in a dry ionomer }

\author{Elshad Allahyarov }

\affiliation{Department of Physics, Case Western Reserve University, Cleveland,
Ohio 44106\\
 and Joint Institute of High Temperatures, Russian Academy of Sciences
(IVTAN), Moscow, 125412 Russia}

\author{Philip L. Taylor}

\affiliation{Department of Physics, Case Western Reserve University, Cleveland,
Ohio 44106}

\begin{abstract}
This simulation study investigates the dependence of the structure of dry Nafion$^{\tiny\textregistered}$-like
ionomers on the electrostatic interactions between the components
of the molecules. In order to speed equilibration, a procedure was adopted which involved detaching the side chains from the backbone and cutting the backbone into segments, and then reassembling the macromolecule by means of a strong imposed attractive force between the cut ends of the backbone, and between the non-ionic ends of the side chains and the midpoints of the backbone segments.  
Parameters varied in this study include the dielectric constant, the free volume, side-chain length, and strength of head-group interactions.  A series of coarse-grained mesoscale simulations
shows the morphlogy to depend sensitively on the ratio of the strength
of the dipole-dipole interactions between the side-chain acidic end
groups to the strength of the other electrostatic components of the
Hamiltonian. Examples of the two differing morphologies proposed by
Gierke and by Gebel emerge from our simulations.
\end{abstract}

\pacs{}
\maketitle

\section{introduction}

Typical proton-conducting polymer electrolyte membranes (PEM) for
fuel cell applications consist of a perfluorinated polymeric backbone
and side chains that are terminated by strongly acidic sulfonic head
groups. They provide the transport medium for the protons generated
during the anodic oxidation reaction of the fuel. PEM fuel cells offer
numerous advantages as power generators for emission-free vehicles,
portable applications, and individual homes \cite{perry2002}. When
such ionomers are exposed to water or humid air, the acidic groups
terminating the side chains dissociate into SO$_{3}^{-}$ groups on
the chains, and protons in the aqueous sub-phase. The sub-phase extends
through the membrane, creating a bi-continuous nano-phase separated
network of aqueous pores and polymer. The percolation structure and
its ability to support proton diffusion have to meet strict requirements
for a PEM to be considered as an `ideal' membrane for fuel cells:
a) It must form a chemically, mechanically, thermally and morphologically
stable structure over a wide range of water content, b) show steady
performance and tolerance of elevated temperatures up to 150$^{\circ}$C
at high proton conductance, and c) be impermeable to gases and contaminating
ions. Development of new PEM membranes that could meet all these demands
is a primary task for many research groups involved in fuel-cell studies.

One extensively used PEM material is Nafion$^{\tiny\textregistered}$,
a perfluorinated polymer from DuPont. The integrity and structural
stability of Nafion membranes is provided by the poly\-tetra\-fluoro\-ethylene
(PTFE) backbones, but it is the hydrophilic clusters of side chain
material that facilitate the transport of ions in the membrane. The
microstructure of these clusters consists of an inter-facial region
of solvated perfluoroether side chains separating the polymer matrix
from the more bulk-like water found in the pores. In the presence
of water, one finds a nano-separation of hydrophobic and hydrophilic
domains. According to the early work of Gierke \textit{et al.} \cite{Gierke1,Gierke2}
these pores are inverted micellar spheres with diameters around 4
nm, and are interconnected by channels with a diameter and length
of about 1 nm.

Recent experimental investigations suggest that the spherical shape
and uniform spacing of these pores are serious oversimplifications
\cite{paddison-review-1}. New small-angle X-ray scattering (SAXS)
and small-angle neutron scattering (SANS) experiments by Gebel \textit{et
al.} \cite{Gebel} for Nafion membranes in different swelling states
have given a different picture for the morphology. Their results suggest
that elongated bundles of hydrophobic backbone material with diameters
of about 4 nm and lengths greater than 100 nm are surrounded by sulfonate
groups and water molecules. In other words, hydrophobic material is
surrounded by hydrophilic domains.

The fact that controversy continues to surround the nature of the
hydrated morphology and the shape of the domains suggests that there
is still room for theoretical investigation and computer simulations
to address this issue. \textit{Ab initio} electronic structure calculations
of polymeric fragments in contact with water \cite{paddison2} and
quantum molecular dynamics studies on model PEM systems \cite{paddison3}
have provided a basis for understanding the energetically favorable
conformations of Nafion chains and the molecular ingredients of the
conduction process \cite{kreuer2004,li2001}. For systems large enough to contain several examples
of hydrophilic inclusions and their interconnecting networks, however,
treatment of the polymer in an \textit{ab initio} manner, i.e. a full
electronic treatment with molecular orbital theory, is not computationally
feasible.

The objective of this paper is to develop a deeper understanding of
the driving mechanism behind the nanophase aggregation of sulfonate
head groups, and to test the extent to which coarse-graining methods
can be applied. We attempt this by using a simple dipolar head group
model for ionomer side chains. We show that the dipole-dipole correlations
strongly depend on the molar fraction of acidic groups and dielectric
permittivity of the membrane. Globular micelles formed in the case
of  zero dipole moment evolve to percolated cluster structures for
a non-zero dipole moment of head groups when an artificial attraction
between the end groups of side chains is introduced. Accounting
for the dipole moment of sulfonate head groups thus changes the conformational
structure of clusters from a Gierke model of spherical aggregations to a
Gebel model of elongated aggregations. The determination of the parameters
necessary for success in modeling nanophase aggregation will help
us understand the essential role played by the electrostatic interactions
of the head group in determining microstructure formation in PEM materials.

The rest of this paper is organized as follows. After a brief review
of segregation models in section II, we present our dipolar model
for side chains in section III.  Section IV is devoted to our simulation results
for dry PEM ionomers. We conclude with a discussion of results and
planned future work in section V.

\section{nanophase segregation models}

Significant experimental and theoretical efforts have been put into
the characterization of the microstructure of dry and swollen Nafion
membranes. Although the fact of microphase segregation and its influence
on the thermodynamic and transport properties of Nafion membranes
are well established, some uncertainty regarding the morphology of
the subphases in swollen Nafion remains. It has been generally assumed
that the hydrophilic phase is continuous, in accord with the experimental
evidence of high conductivity and water permeability. The conventional
model of microphase segregation in Nafion membranes, put forward by
Gierke \textit{et al.} \cite{Gierke1,Gierke2}, where the protons and water
diffuse between globular clusters through cylindrical narrow channels,
was further modified by Mauritz and Rogers \cite{mauritz} and
Eikerling \textit{et al.} \cite{eikerlinz}. Also, Ioselevich \textit{et al.} \cite{ioselevich} 
developed a lattice based
micelle-channel model to describe
a connecting structure of a water bridges between the cages of the
micelles. However, there is as yet no direct experimental evidence for the existence
of the channels connecting the clusters in Gierke's model. Other models
explaining the properties of hydrated Nafion exploit bilayer, lamellar,
and sandwich-like structures for nanophase separation in PEM membranes.
Gebel \cite{Gebel} describes the Nafion membrane as an aggregation
of polymeric chains forming elongated objects (simplified as cylinders),
embedded in a continuous ionic medium. At larger scales, those aggregates
form bundles with definite orientational order. This new picture of
multi-scale structure can explain membrane swelling as a continuous
process from the dry state to a colloidal suspension.

Scanning calorimetry experiments on cast Nafion films \cite{zanderighi}
led to the conclusion that a continuous water phase exists only when
the relative humidity of the membrane exceeds 90\%. This conclusion
contradicts the simulation results of Khalatur \textit{et al.} \cite{khalatur},
where  formation of continuous channels at very low hydration, and
even in almost dry membranes, is reported. This is certainly consistent
with the conductivity measurements in Nafion films with very little
water present \cite{zawodzinski1991}. In contrast to all the above-mentioned
network models, Vishnyakov \textit{et al.} \cite{vishnyakov2001}
showed that the water clusters do not form a continuous hydrophilic
subphase. Instead, the cluster-size distribution evolves in time as
a consequence of the formation and rupture of temporary bridges between
clusters.

Molecular dynamics (MD) studies, using both classical force fields
and an \textit{ab initio} approach, appear to support the idea of
irregularly shaped ionic clusters of sulfonic groups connected by
smaller channels of similar composition embedded in the hydrophobic
covalent matrix. However, the characteristic length scales of the
morphological conformations suggested by experimental data are currently
inaccessible to atomistic modeling. To probe the morphology at the
nanometer scale, a mesoscale model is required.

\section{Dipole model for side chains. }

In this work we are concentrating on the effects of electrostatic
interactions among the elements of the side chains, and so in order
to avoid the excessively slow processes resulting from chain entanglement
we consider a segmented system in which the side chains are detached
from the backbone. The later introduction of very strong interactions
between the ends of the side chains and the backbone, which have the
effect of knitting together the segments into a single macromolecule,
do not seem to affect our results significantly.

A generic acidic side chain of Nafion is commonly denoted as CF$_{3}$OCF$_{2}$CF(CF$_{3}$)O(CF$_{2}$)$_{2}$SO$_{3}$H.
In order to expedite these simulations, we adopt a simplified model
for the side chain by using a united-atom presentation of the CF$_{2}$
and CF$_{3}$ groups \cite{wescott2006,vishnyakov2001,yamamoto} and the SO$_{3}^{-}$
sulfonate head groups \cite{spohr2004}. These groups are modeled
as Lennard-Jones (LJ) monomers. We employ the same united-atom approach
for backbone chain segments.

In the framework of our dipole model all the partial charges on the
ether oxygen, carbon and fluorine atoms of the side chain are set
to zero. The fluorocarbon groups of the backbone skeleton are also
assigned zero partial charges. The electrostatic charge is located
entirely on the sulfonate head group, SO$_{3}^{-}$,
and on the hydrogen ion H$^{+}$ permanently bound to it. The dipole moment produced when the proton
and the sulfonate group, modeled as LJ monomers, carry the full formal
charges of $+e$ and $-e$ respectively, was reduced by a factor $D_r$ in some of our simulations.  Though our model is not applicable for the measurement of
proton diffusion in PEM membranes, it is a good starting point for
a step-by-step exploration of the nanophase morphology in dry PEM
materials.

The overall inter-monomer interaction in our model system is a combination
of electrostatic Coulomb interactions between each pair of charged
particles, 12-6 Lennard-Jones (LJ) potentials between neighboring
monomers, and intrachain stretch and bend energies. The Lennard-Jones
interaction parameters for the side chain are chosen to agree in most
instances with the Nafion model of Paddison \cite{Paddison1} . We
make a further simplification for the LJ parameters of the side chain
by assigning  a universal value of $\sigma$, chosen to be 3.5 \AA~     for all monomers. The
total potential energy of the system is then given as \begin{equation}
E_{{\textrm{total}}}=E_{LJ}+E_{Q}+E_{{\textrm{bond}}}+E_{{\textrm{angle}}},\end{equation}
 where $E_{LJ},\, E_{Q},\, E_{{\textrm{bond}}},$ and $E_{{\textrm{angle}}}$
are the Lennard-Jones, electrostatic, bond-stretching (two-body term),
and angle-bending (three-body term) components of the total energy,
and are given by \begin{equation}
E_{LJ}(r)=4\varepsilon_{LJ}(\sigma^{12}/r^{12}-\sigma^{6}/r^{6})\end{equation}
\begin{equation}
E_{Q}=\sum_{i>j}\frac{Q_{i}Q_{j}e^{2}}{\epsilon R_{ij}}\label{eq:2}\end{equation}
\begin{equation}
E_{{\textrm{bond}}}(R)=\frac{1}{2}k_{b}(R-R_{0})^{2}\label{eq:3}\end{equation}
\begin{equation}
E_{\textrm{angle}}(\theta)=\frac{1}{2}k_{\theta}(\theta-\theta_{0})^{2}\label{eq:4}\end{equation}
 We omit consideration of the dihedral potential term (four-body term),
as this would be expected to make only a small contribution to the energetics
of the single chain conformations in these dense polymeric systems.  (This assumption was tested in sample simulations in which this term was restored, and found to have little effect on our results.)
The following parameters for the polymer chains were used: an equilibrium
bending angle $\theta_{0}=110^{\circ}$, an equilibrium bond length
$R_{0}=1.54 \AA$ (0.44$\sigma$), a bending force constant $k_{\theta}=120$ kcal/mol
deg$^{2}$, and a stretching force constant $k_{b}=700$ kcal/mol
$\AA^{2}$. The last two constants correspond to the thermal energy
$k_{B}T$ at room temperature for a bond-length displacement of 0.1
\AA \, and bond-angle displacement of $1^{\circ}$ from the equilibrium
position. The bond length between the SO$_{3}^{-}$ head-group monomer
and the hydrogen ion was chosen to be $\sigma$, and hence 0.35 nm. The LJ interaction coefficient
$\varepsilon_{LJ}$ was chosen to be 3$k_{B}T$ for hydrophobic-hydrophobic
(HH) interactions, and 1$k_{B}T$ for both hydrophobic-hydrophilic
(HP) and hydrophilic-hydrophilic (PP) interactions. In the latter
case The LJ potential was modified  to be purely repulsive by truncating at its minimum, where $r = 1.1224\sigma$, and raising it by the addition of an amount $\varepsilon_{LJ}$.   The reason for this choice is that the interaction between protons is not well represented by van der Waals forces, which arise from virtual dipole-dipole interactions, and also that such attractive forces are small in comparison to the Coulomb interaction.
 The normalized charge $Q_l$ ($l=i,j$) in Eq.(3)  is +1 for protons and -1 for sulfonate
groups.

We used this coarse-grained model to perform molecular dynamics simulations
of ensembles of ionomer segments.   Because chain entanglement greatly slows the rate of equilibration in polymers, we adopted the strategy of temporarily cutting the system into more mobile segments, which were later reassembled.   In order to increase the diffusional
rate of our model ionomer we detached the side chains from
the backbone skeleton, an approach previously adopted by Vishnyakov
\cite{rivin-vishnyakov}, and also cut the backbone skeleton into
segments \cite{glotzer2002}.  In total the system then  consists of $N_s$ sidechain segments
and $N_b$ backbone segments.   Taking into account that the sulfonic
acid groups are hydrophilic while the ethers and fluorocarbon groups
are hydrophobic \cite{Paddison1}, we adopted the following general
representation for a coarse-grained architecture of side chains and
backbone segments. Each side chain contains $N_m=N_P + N_H$ monomers
of which $N_{P}$ are hydrophilic and $N_{H}$ are
hydrophobic. Its architecture is written as: $N_{H}$H+$N_{P}$P.
The backbone segments are fully hydrophobic with an architecture $n$H.
In the majority of systems studied in this paper,  the side chain architecture was 8H+2P and the backbone segment architecture was 10H.

The simulations were then performed in several stages. The initial configuration was first prepared by using Monte-Carlo
techniques to grow a mixture of polymer chain segments  with fixed bending
angle $\theta_{0}$ and bond length $R_{0}$.   After an initial equilibration the constraints on $\theta_{0}$ and $R_{0}$ were removed.  Subsequently, MD runs
were performed for time periods between 50ps and 500ps at constant
volume $V$ and constant temperature $T$ until the system of polymer
segments was fully equilibrated. The system temperature was controlled
by coupling the system to a Langevin thermostat with a friction coefficient
$\gamma=0.1$ and Gaussian white-noise force of strength $6k_{B}T\gamma$.
The equations of motion were integrated using the velocity Verlet
algorithm with a time step of 0.5 fs. We imposed standard periodic
boundary conditions to our system, filling space with translational
replications of a fundamental cell. Long-range electrostatic interactions
were calculated using the standard Lekner summation algorithm \cite{Lekner}.

To check whether equilibrium had been reached we examined the proton
pair distribution function $g(r)$ in every picosecond time interval,
and continued the run until no significant further change was observed.
To avoid mistaking the trapping of the simulated system into a metastable
glassy state for true equilibrium, we repeated each run with several
different initial configurations. For each computer experiment we
typically gathered system statistics over time intervals of a few
nanoseconds. We believe this to be sufficient to allow for the self-organization
of the side chains and backbones into clustered structures.

 In the next stage of the simulations the segments were reassembled
 into the branched chain characterizing the original Nafion. 
This was achieved by the simultaneous  introduction of  a specific
mutually attractive force between the ends of each backbone segment,
which united them into a single chain, and a similar force acting
between the tail monomer of the detached side chains and the fifth
monomer of every backbone segment.  This was applied in the form of a
LJ potential of strength $\varepsilon_{LJ} = 6 k_BT$. 
To avoid the formation of  star-like branched
polymers,  only a single occupancy of the backbone attachment sites was permitted.
The simulation was then resumed, and run until a new equilibrium was attained.
It was found that the bonds on the ends of all the backbone segments
were occupied, and that a large majority (typically around 95\%) of
the side chains were reattached to a backbone segment. This fraction
was a function of the ratio $\psi$ of the number $N_s$ of side chains
(and hence of sulfonate groups) to the number  $n \times N_b$ of
backbone monomers.   The quantity $\psi$ thus measures the
concentration of sulfonate groups relative to the concentration of
backbone monomers.  The few side-chains that did not bond to the special sites on the backbone attached themselves more loosely to the backbone through the weaker regular hydrophobic-hydrophobic LJ interaction  potential.

We also introduce the absolute molar concentration of sulfonate groups,
 $\eta \equiv N_s/N_0 V_{\rm cell}$, with $N_0$ being Avogadro's
 number and $V_{\rm cell}$ the volume $L^3$ of the simulation cell, as
 an alternative parameter with which to describe the sulfonate
 concentration, independently of the amount of backbone polymer.  
 While it is true that in real dry Nafion, no significant free volume
 exists, and so the  total number of monomers per unit volume is
 roughly fixed at the time of synthesis, in a simulation, however, it
 is possible to change the density of the material.  It is of interest
 in understanding the mobility of ions in Nafion to know what
 structures would form in the presence of imposed free volume, were
 such a reduction in density of the material possible.

In order to observe the effects of  the dipolar nature of
the sulfonate-proton group on the structural characteristics
of side chain aggregates, we calculated the normalized pair distribution function
$g(r)$ of head group protons (which are the end group monomers of
the side chains). The function $g(r)$ indicates the relative probability
of finding two protons at a separation distance $r$ averaged over
the equilibrium trajectory of the simulated system, so that
\begin{equation}
g(r)=\left(V_{\rm{cell}}/N_{+}\right) \frac{dn_{+}(r)}{4\pi r^{2}dr}.
\end{equation}
 Here $d n_{+}$ is the number of protons located in a shell of thickness
$dr$ at the distance $r$ from a fixed proton, and $N_{+}$ is the
total number of protons in the system of volume $V_{\textrm{cell}}$.

In addition to this spherically averaged quantity, it is also useful
to note some indications of the local anisotropy in order to distinguish
between the Gierke and Gebel models. To this end we also evaluated
the partial structure factors \[
S(q_{j})=\left\langle \left[\sum_{i=1}^{N_{+}}\cos q_{j}r_{i}\right]^{2}+\left[\sum_{i=1}^{N_{+}}\sin q_{j}r_{i}\right]^{2}\right\rangle .\]
 Here $j$ denotes the $x$, $y$ and $z$ components of the vector
${\bf q}$, with $q_{j}=2\pi n_{j}/L$, and the sums can proceed
over the entire sample or over smaller volumes.

\begin{table} [!ht]

\caption{List of key variables.}
\begin{ruledtabular}
\begin{tabular}{lc} 


$D_r$ & dipole moment reduction factor  \\ 
$\theta_0$ & equilibrium bending angle for backbone and sidechain polymers \\
$R_0$ &  equilibrium bond length for backbone and sidechain polymers \\
$ k_{\theta} $ & bending force constant \\
$ k_b $ & stretching force constant \\
$ k_B $ & Boltzmann constant \\
$T$ & system temperature \\ 
$\sigma$ & monomer diameter \\
$\varepsilon_{LJ}$ &  Lennard-Johns interaction parameter between monomers\\
$Q_e$ & normalized charge of sulfanates and ptorons \\ 
$\epsilon$ & dielectric constant of medium \\ 
$N_s$ & number of side chains \\ 
$N_b$ & number of backbone chain segments \\ 
$N_m$ & number of monomers per side chain \\ 
$N_H$ & number of hydrophobic monomers per side chain \\ 
$N_P$ & number of hydrophilic monomers per side chain \\ 
$n_b$ & number of hydrophobic monomers per backbone chain segments \\ 
$\psi$ &  ratio of the number of sulfonate groups to the number of backbone monomers.\\
$\eta$ & molar concentration of head groups \\ 
$N_0$ &  Avogadro's number \\ 
$L$ & length of simulation box \\
$g(r)$ & proton-proton pair distribution function \\ 
$n_+$ & local number  of protons \\ 
$N_+$ & number of protons in the system \\ 
$V_{cell}$ & volume of simulation box \\ 
$S(q_j), \, j=x,y,z$ & partial structure factors \\ 
$\Gamma$ & coupling parameter in ionomer \\ 
$\lambda_B$ & Bjerrum length \\ 

\end{tabular}
\end{ruledtabular} 

\label{tab1} 

\end{table}

\begin{table}[!ht]

\caption{\label{tableI} Parameters used in  combinations for
  simulation runs. Here $\eta$ is the molar sulfonate-ion concentration, $N_b$ the
  number of backbone chain segments,  $L/\sigma$ the simulation cell size in
  units of $\sigma$ = 0.35 nm, $N_m$ is the number of monomers per sidechain, $\psi$ is ratio of the
  number of sulfonate groups to the number of backbone monomers, and
  $\epsilon$ is the dielectric constant of the medium.    }
\begin{ruledtabular}
\begin{tabular}[t]{lccccccc}

Runs   &  $\eta \textrm{(mol/l)}$  &
$N_b$   &  $L/\sigma$  &  $N_m$ & $n_b \psi$ & $\epsilon$ & $D_r$  \\ 

\colrule 

Run 1   &   0.8      &  0     &  30  &  10  & 0 & 1, 15, 80 & 1  \\  
Run 2   &   0.8      &  500   &  30  &  10  & 1 & 1, 15, 80 & 1  \\ 
Run 3   &   0.8      &  1000  &  30  &  10  & 2 & 1         & 1  \\ 
Run 4   &   0.8      &  1500  &  30  &  10  & 3 & 1         & 1  \\ 
Run 5   &   0.8      &  2000  &  30  &  10  & 4 & 1         & 1  \\ 
Run 6   &   0.8      &  2500  &  30  &  10  & 5 & 1         & 1  \\ 
Run 7   &   1.6      &  0     &  30  &  10  & 0 & 1, 15, 80 & 0, 0.5,  1  \\ 
Run 8   &   1.6      &  1000  &  30  &  10  & 1 & 1         & 0, 0.25, 0.5, 1\\ 
Run 9   &   1.6      &  2000  &  30  &  10  & 2 & 1         & 0, 0.5,  1 \\ 
Run 10  &   0.2      &  1000  &  60  &  10  & 1 & 1         & 1  \\  
Run 11  &   0.5      &  1000  &  45  &  10  & 1 & 1         & 1  \\
Run 12  &   2.5      &  1000  &  25  &  10  & 1 & 1         & 1  \\
Run 13  & 0.16--4.8  &     0  &  30  &  10  & 0 & 1         & 1 \\ 
Run 14  &   1.6      &     0  &  30  &2--10 & 0 & 1         & 1\\

\end{tabular}
\end{ruledtabular}
\end{table}

\section{Simulation results}

The goal of these extensive coarse-grained molecular-dynamics simulations
was to investigate the relative importance of the various model parameters on the aggregation of the sulfonate (acidic) head
groups.     To achieve this we performed a series of simulation runs in which the key parameters were varied as summarized in Table II.   These parameters included
the absolute concentration and the relative concentrations of side chains and backbone polymer, as well as the strength of the electrostatic interactions.   This last quantity is 
characterized
by a Coulomb coupling parameter defined as $\Gamma=e^{2}/\epsilon k_{B}T\sigma \equiv \lambda_{B}/\sigma$.
Here  $\lambda_{B}$, which is known as the Bjerrum length, is about  56~nm  for a dielectric constant $\epsilon=1$
and $T=300$ K. The coupling parameter $\Gamma$ was varied between
2 and 160 by varying the system dielectric permeability $\epsilon$
between 80 and 1, with higher $\Gamma$ representing stronger electrostatic
correlations in the system. In a typical simulation, a few thousand
side chains are confined in a cubic box whose side is of length $L=30\sigma$, with
$\sigma=0.3$ nm being the monomer diameter.


Each simulation run was started from a randomly generated assembly
of side chains and backbone segments within the system volume. However,
in order to avoid packing problems, and to make this process effective for high monomer
densities, a preliminary simulation of the same number of monomers was first performed
in a larger box of side $L'>L$, and this  was then followed by a gradual reduction of the
system volume until the desired molar concentration was achieved.

\subsection{The role of the strength of the electrostatic interactions}

\begin{figure}[h]

\caption{(Color online) Proton-proton pair distribution function $g(r)$ for (a)
Run~1 (low density), and (b) Run~7 (high density) for $\epsilon$=1 (dot-dashed
line), 15 (dashed line) and 80 (full line). Simulation results for
$\epsilon$=1 are scaled down by a factor of four to fit the figure dimensions.
The $\epsilon=1(*)$ (symbols) in (b), given for comparison, corresponds to the $\epsilon=1$ line
in (a). \label{fig4}}
\includegraphics*[width=0.6\textwidth]{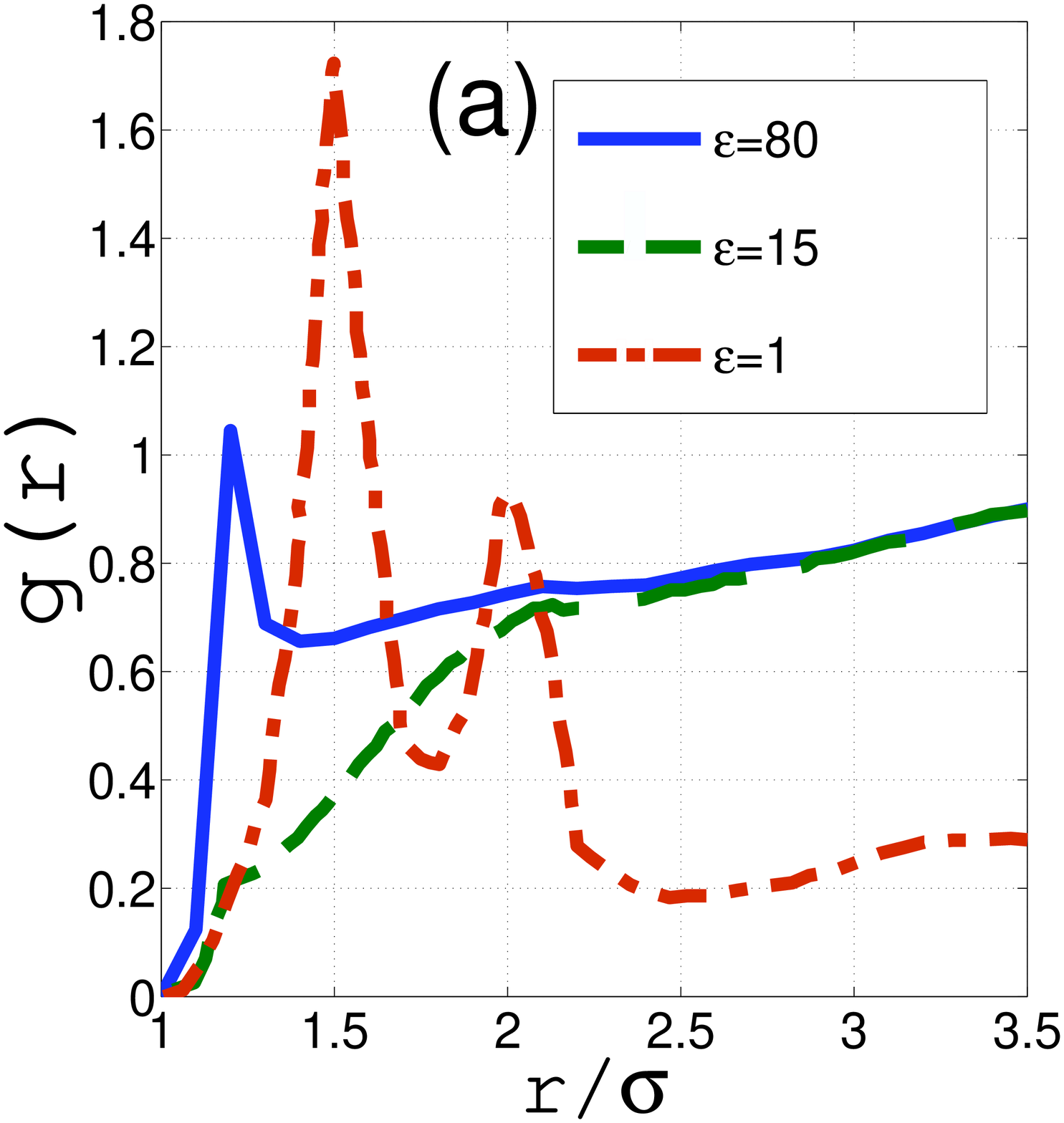} 
\includegraphics*[width=0.6\textwidth]{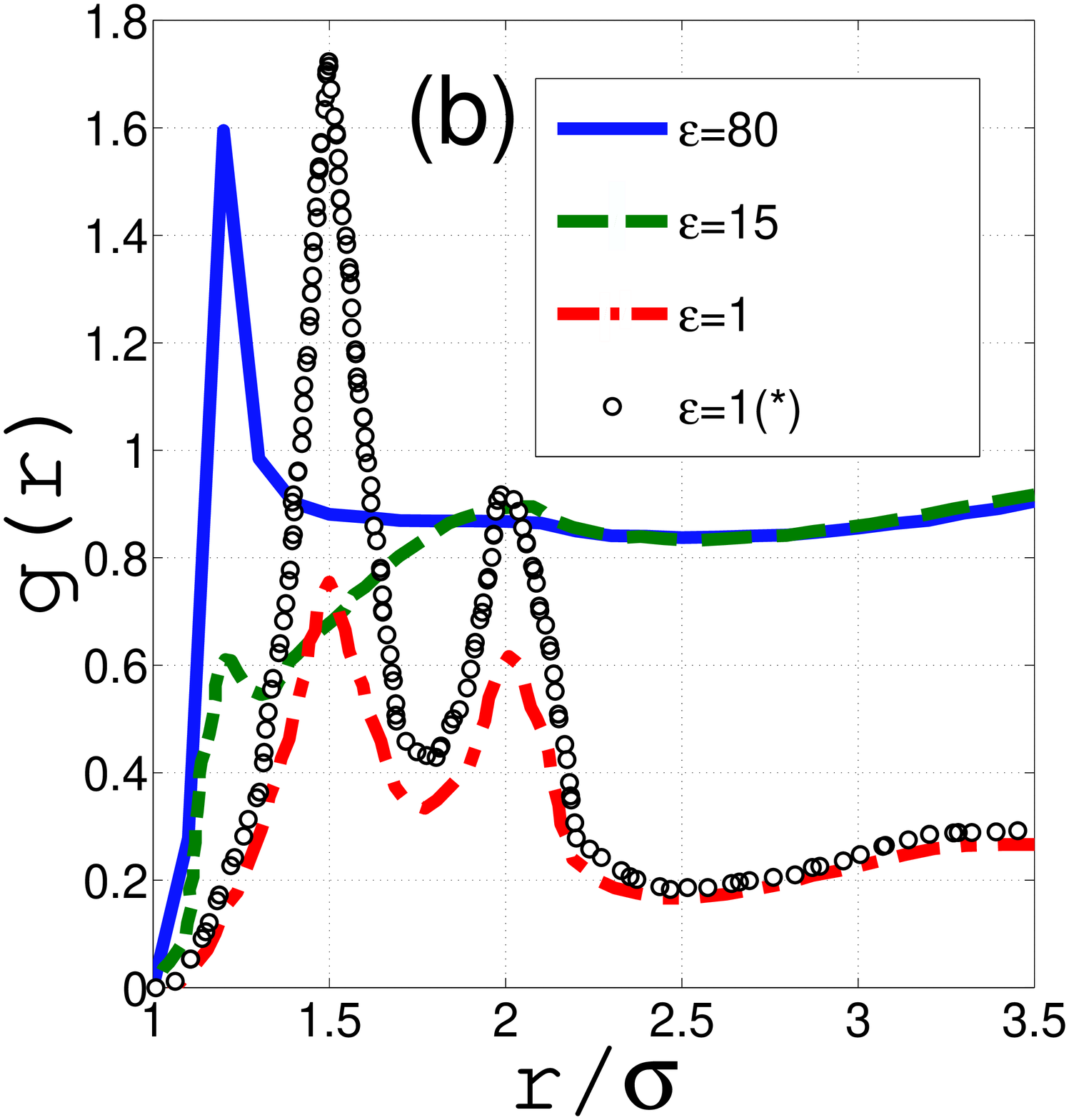}

\end{figure}

\begin{figure}[h]

\caption{(Color online) Simple multiplet structures in side chain aggregates.
Upper row: compact (left ) and linear (right) multiplets. Bottom row:
branched multiplet. The hydrophobic tails are shown only for a compact
multiplet.  The dashed objects C and D for the branched cluster indicate
another possible configuration of objects A and B respectively. Positively charged hydrogen atoms are
shown as yellow circles. \label{clusters}}
\includegraphics*[width=0.8\textwidth]{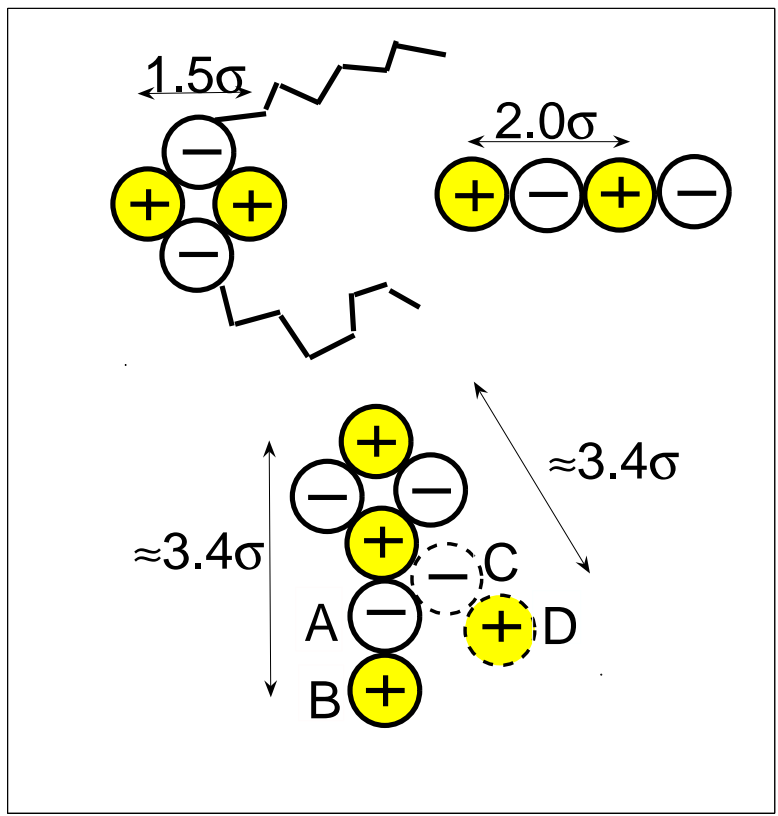}

\end{figure}

In this first simulation the constraints imposed by the attachment of the side chains to the backbone were removed in order to concentrate on the effects of the electrostatic interactions.
We then analyzed the proton-proton pair distribution function
$g(r)$ calculated for an assembly of unattached side chains for three different dielectric constants $\epsilon$
of the ionomer medium, and for two different molar concentrations.   The higher concentration, for which $\eta = 1.6$ mol/l, is characteristic of typical dry Nafion samples.   Figure \ref{fig4} shows the proton-proton pair distribution function for the parameters described in Table II as
Run~1 and Run~7. The first observation is that, as expected, small $\epsilon$,
or, equivalently, high $\Gamma$, facilitates clustering of the head
groups, as shown in the red dot-dashed lines in Figure \ref{fig4}. The
graph of $g(r)$ shows well-defined maxima at $r/\sigma=1.5$,\,2.0,
and 3.4. A detailed investigation shows that this is a consequence
of the formation of three simple multiplet structures: compact and
linear multiplets, each created by two head groups, and a branched
multiplet composed of three head groups, as shown in Fig.~\ref{clusters}.

For $\epsilon=80$ there is no correlation between head groups on
different side chains except for steric hindrances, as seen in the
continuous blue lines in Figure \ref{fig4}. A more surprising result
is the effective repulsion between acidic head groups at intermediate
dielectric permittivities, as seen in the green dashed lines in
Fig.~\ref{fig4}. It appears that, in the absence of strong correlations,
each head-group dipole prefers to stay in the bulk of the sample in
order to increase its polarization energy. This is similar to the
well known salt-depletion effect in colloidal physics, when salt pairs
are effectively repelled from the surface of a neutral colloid \cite{protein_paper}.

The other unexpected result is the suppression of the correlations
between the end groups at high molar sulfonate concentrations $\eta$, as seen in
the dot-dashed line and symbols in Figure \ref{fig4}(b).
This can be traced back either to the reduction of the Debye screening
length, which effectively shields the dipoles from each other, or
to the enhancement of the hydrophobic-hydrophobic attraction in the
system. The latter emerges from the hydrophobic tails of the
side chain segments. 

\begin{figure}[h]

\caption{(Color online) (a) Proton-proton pair distribution function
$g(r)$ for Run~2 and different dielectric permittivities. $\epsilon$=1
(dot-dashed line), 15 (dashed line) and 80 (full line). Results for
  Run~1, not shown here, are very close to the results for  $\epsilon$=1.  
 Simulation results for $\epsilon$=1 are scaled down by a factor of
 four to fit the figure dimensions.  (b)  Proton-proton pair
 distribution function $g(r)$ for different amounts of backbone
 polymer (Runs 2 and 6).  Results for Run 3-5 (not shown here) that correspond
 to intermediate amount of backbone polymer, lay between these two
 curves. \label{fig6} } 
\includegraphics*[width=0.6\textwidth]{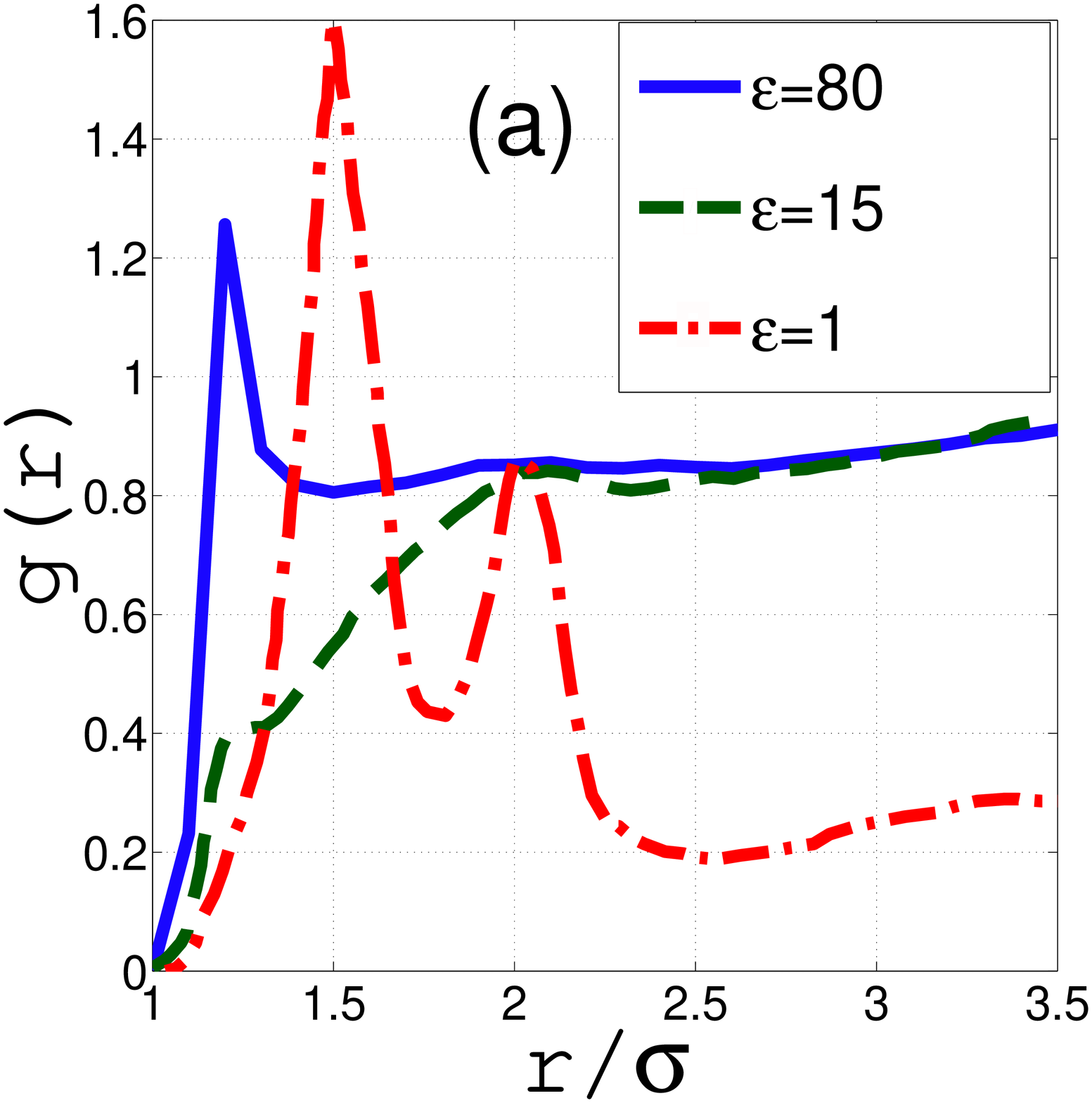}
\includegraphics*[width=0.6\textwidth]{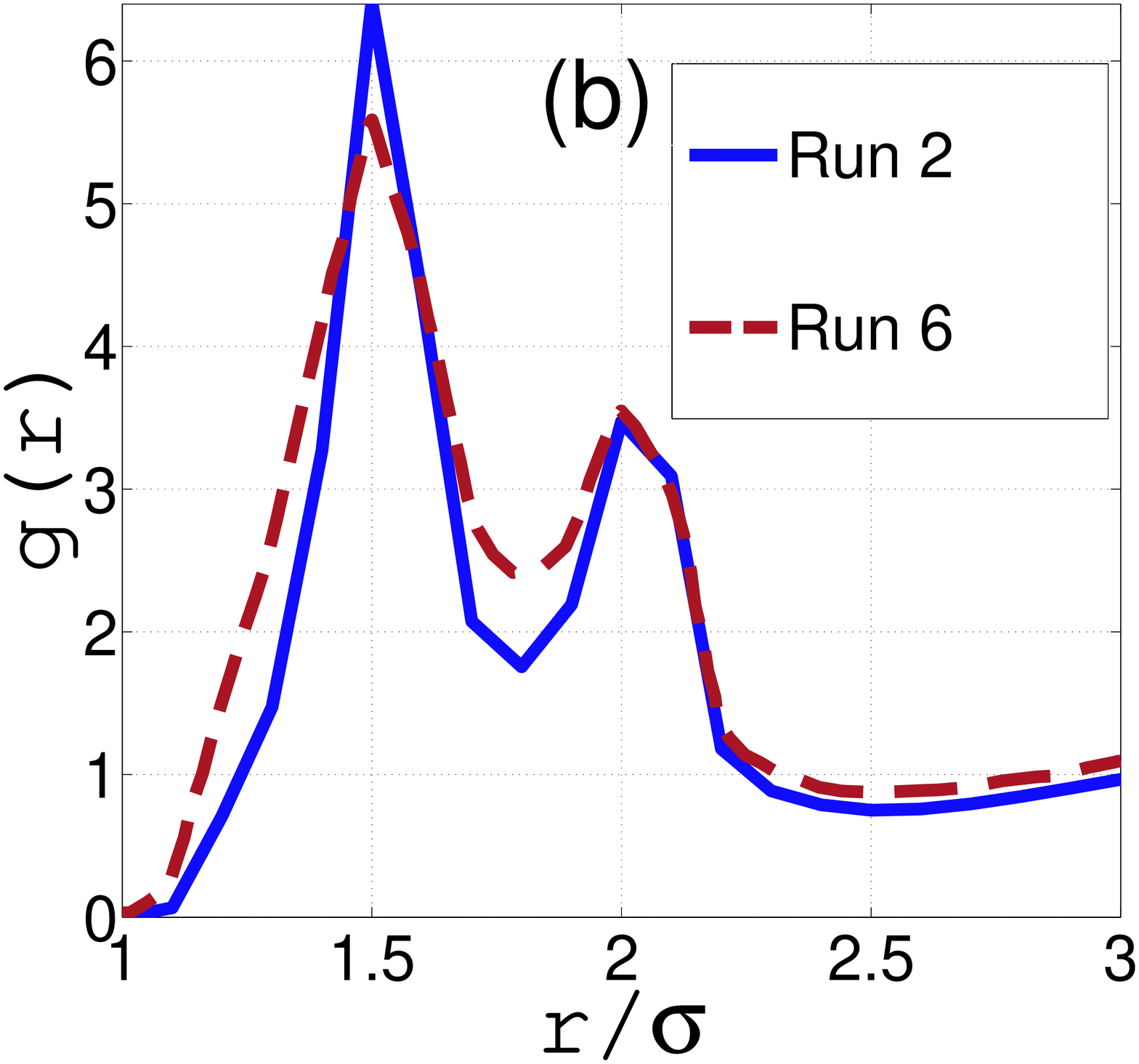} 

\end{figure}

To decide which of these two possible mechanisms was responsible, we added into the system
described in Fig.~\ref{fig4}(b) a small number of backbone segments
in order to boost the effective hydrophobicity (or association energy)
in our model system.   One backbone segment of 10 monomers was added
for each side chain unit, making $\psi = 0.1$.   After equilibration,
the side chains were then loosely attached to the backbone segments
and the backbone segments were mostly united in the manner previously
described, whereupon the system was further equilibrated. The results,
shown as Run~2 in Figure \ref{fig6}, 
clearly indicate that the backbone segments have a minor impact on
the head group distribution function. The system has almost the same
microphase structure as observed for the original side chain system.
A fivefold increase in the backbone concentration in Figure \ref{fig6}(b) has only a small
effect on $g(r)$. Therefore, it is the dipole-dipole interaction
that tends to destroy the correlations when the molar fraction of
sulfonates is high. This effect takes place when the screening length
becomes less than the mean separation distance between the dipoles.
This has some relevance to the search for ionomer membranes with superior
transport properties: whereas for better proton conductance one needs
a higher sulfonate molar fraction, the latter hinders sulfonate aggregation,
which is vital for water-sulfur channel formation in wet ionomers.

\begin{figure}[h]

\caption{(Color online) Typical snapshots of a system for Run~4 (case (a)),
and Run~6 (case (b)). Spheres represent the end-groups (protons)
of side chains. The sulfonate group and neutral chain sections are
drawn by lines (in red in online version). Different bead colors correspond
to different bead altitudes, with a blue color for low-altitude beads
(at the bottom of simulation box) and a red color for high-altitude
beads (at the top of simulation box). The size of all the structural
elements is schematic rather than space filling. \label{fig5}}
\includegraphics*[width=0.65\textwidth]{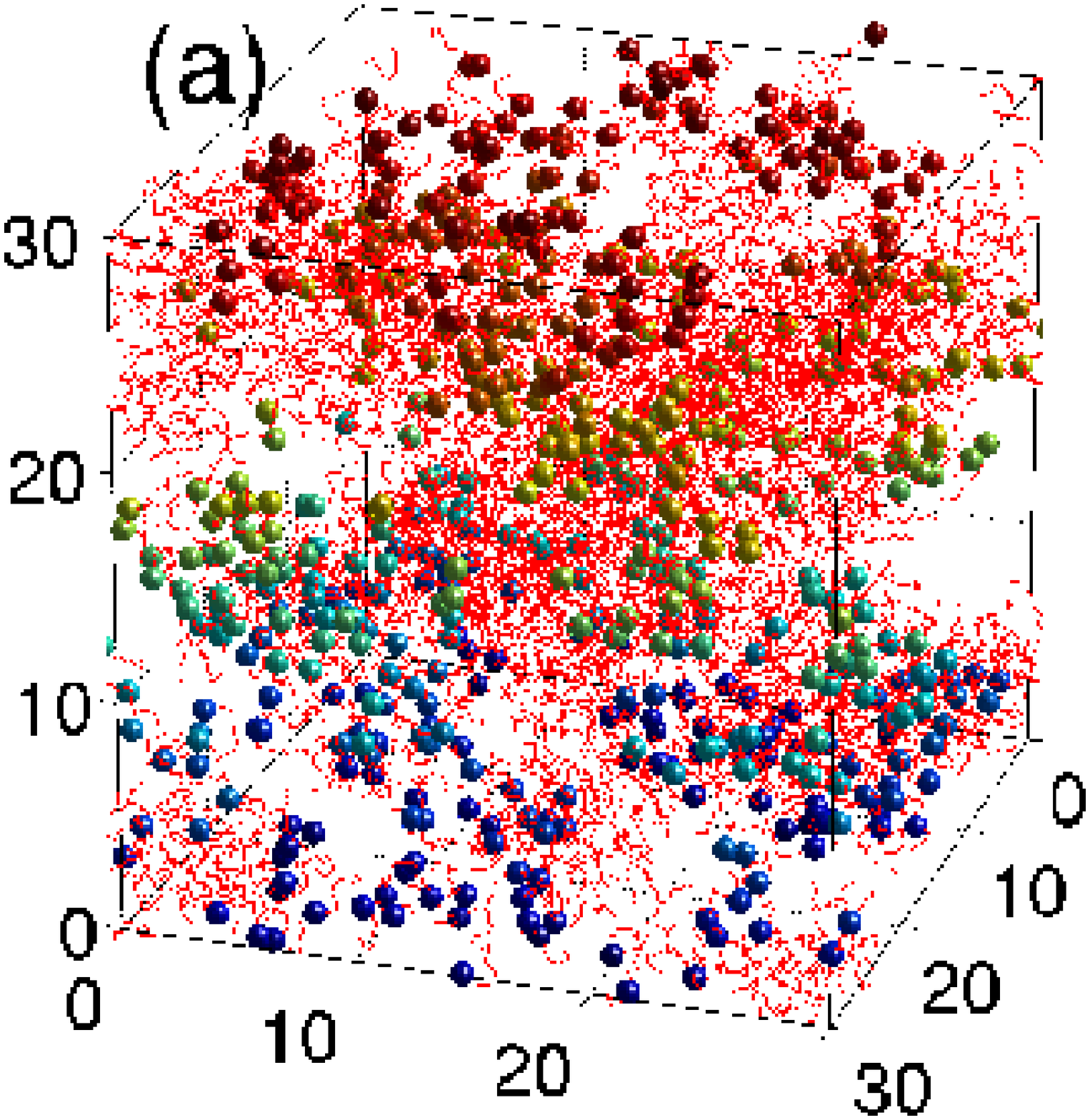}
\includegraphics*[width=0.65\textwidth]{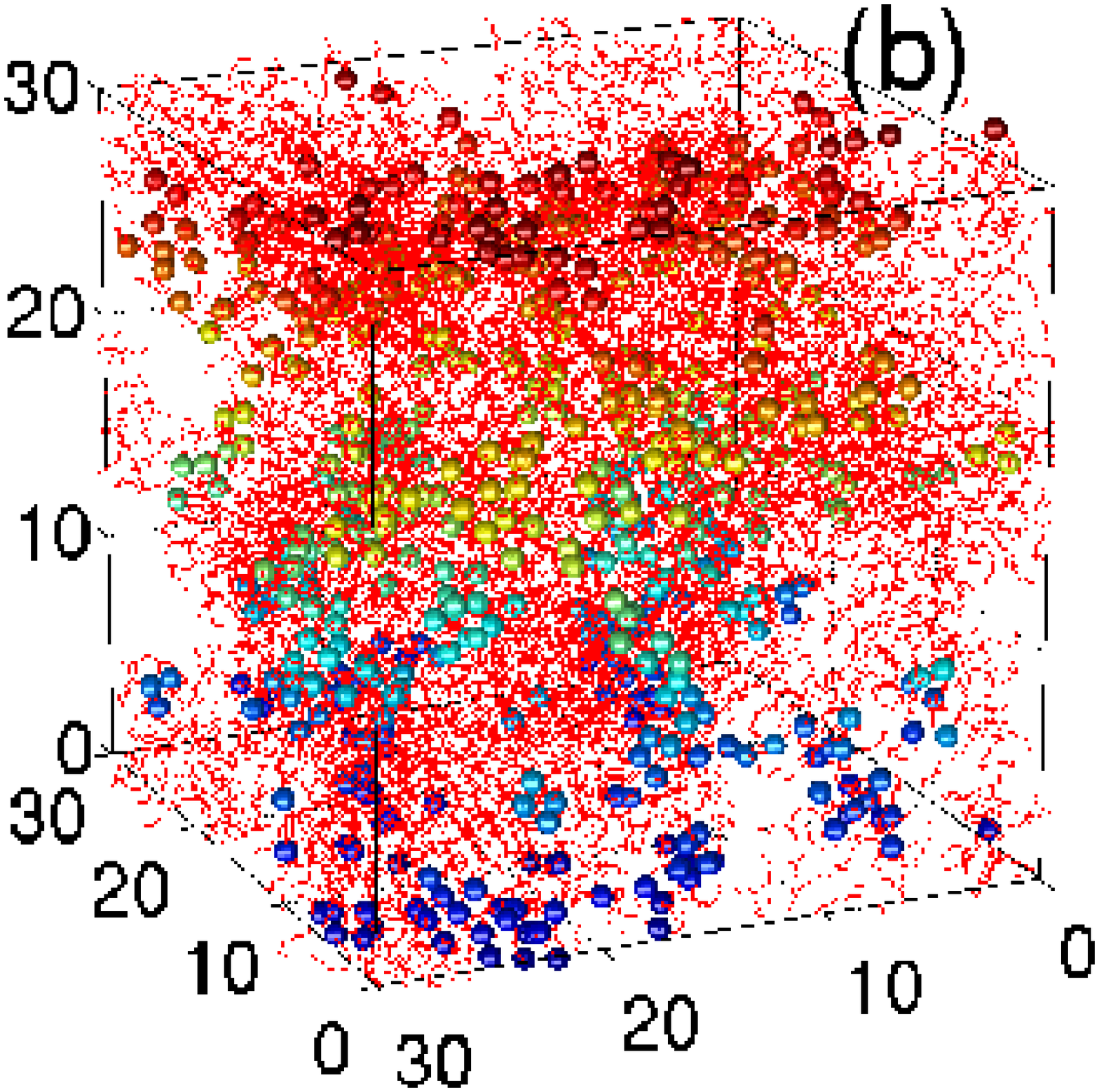} 
\end{figure}

Typical snapshots of the simulation box, shown in Figure \ref{fig5},
make visible the process of microphase separation in ionomer systems.
The tips of the side chains, which correspond to end group protons,
are shown as balls, whereas the rest of the side chains are plotted
as lines. The backbone segments, drawn also as lines, mostly cluster
together to create large aggregates. The side chains either anchor
onto the surface of the backbone aggregate, or create their own small
clusters.

\begin{figure}[h]

\caption{(Color online) A short-time average of the 3D density of monomers
  for Run~5. The hydrophobic tails of side chains 
and backbone segments create a solid matrix, whereas the hydrophilic
head groups forms the walls of the channels. \label{fig3d-density}}
\includegraphics*[width=0.9\textwidth]{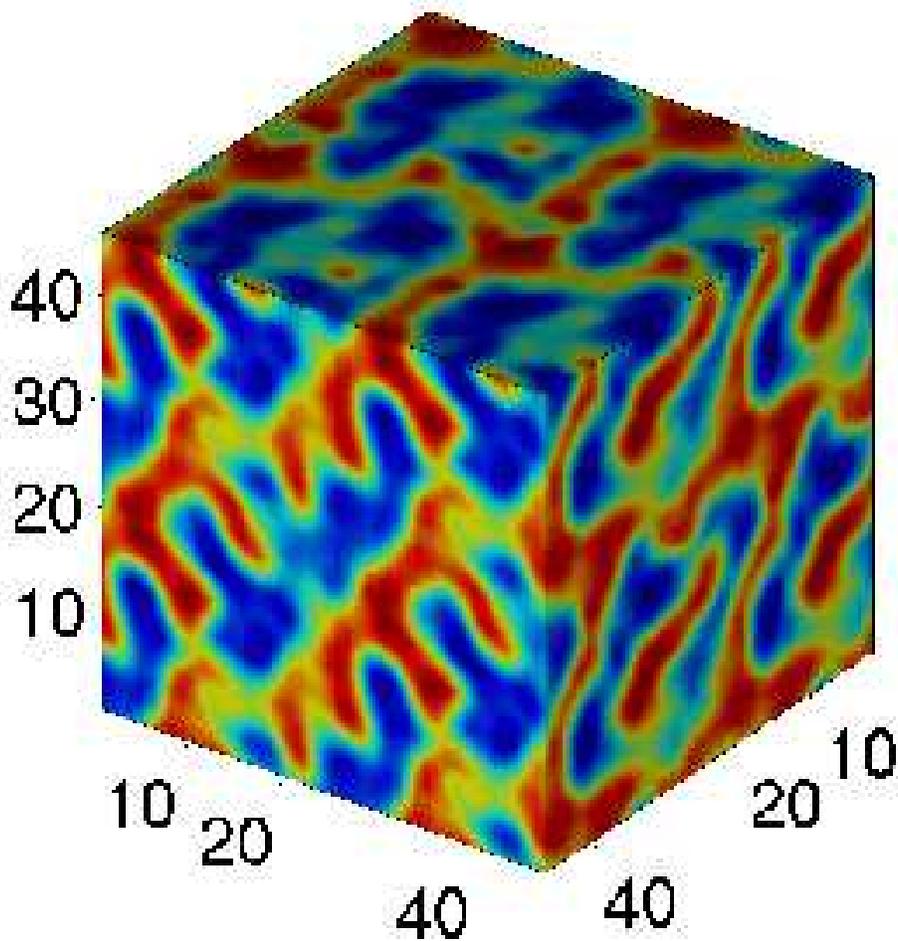} 

\end{figure}

Figure \ref{fig3d-density} presents a 3-dimensional density distribution of
monomers for Run~5. The areas of low (high) density colored in blue (red) show the
channel structure of the aggregates. The polymer matrix is composed
of hydrophobic tails of side chains and backbone segments. The hydrophilic
head groups form the walls of these channels. Though this 3D density
picture differs from the long-bundle structural model of Nafion \cite{Gebel},
it supports the model of a dynamic system of non-spherical clusters
with temporary inter-cluster bridges \cite{vishnyakov2001}.

\subsection{Structure dependence on sulfonate molar concentration $\eta$.}

\begin{figure}[h]

\caption{(Color online) Proton-proton pair distribution function $g(r)$ for
different box sizes $L$, and hence sulfonate concentrations $\eta$, for Runs 8 and 10-12. The first peak of $g(r)$
for Run~10 (full line, blue in colored figure) has a value $g(r=1.5\sigma)=37$,
out of the scale of the plot. \label{fig9}}
\includegraphics*[width=0.9\textwidth]{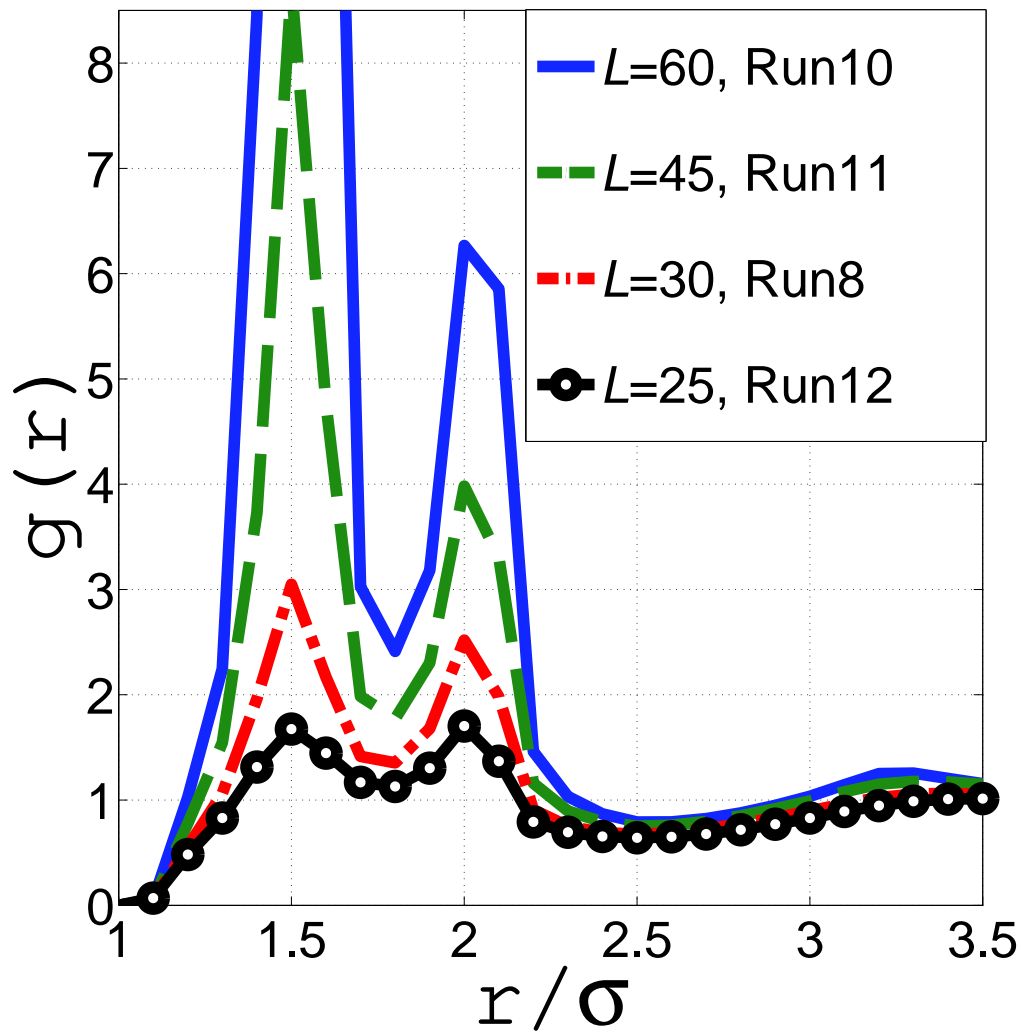} 
\end{figure}

\begin{figure}[h]

\caption{(Color online) Proton-proton pair distribution function $g(r)$ for
different sulfonate molar concentrations $\eta$, Run~13. (a):
0.16 mol/l $\le \eta \le$ 1.6 mol/l. (b): 1.6 mol/l $\le \eta \le$
4.8 mol/l. The case $\eta=1.6(*)$ (line with circles in (b)) is for
Run~9. The first peak of $g(r)$ for $\eta=0.16$ ($\eta=0.32$) mol/l 
(full blue line (dashed green line) in (a)) has a value
$g(r=1.5 \sigma)=53$ ($g(r=1.5 \sigma)=22$), which is out of the scale
of the plot. \label{fig9a}} 
\includegraphics*[width=0.65\textwidth]{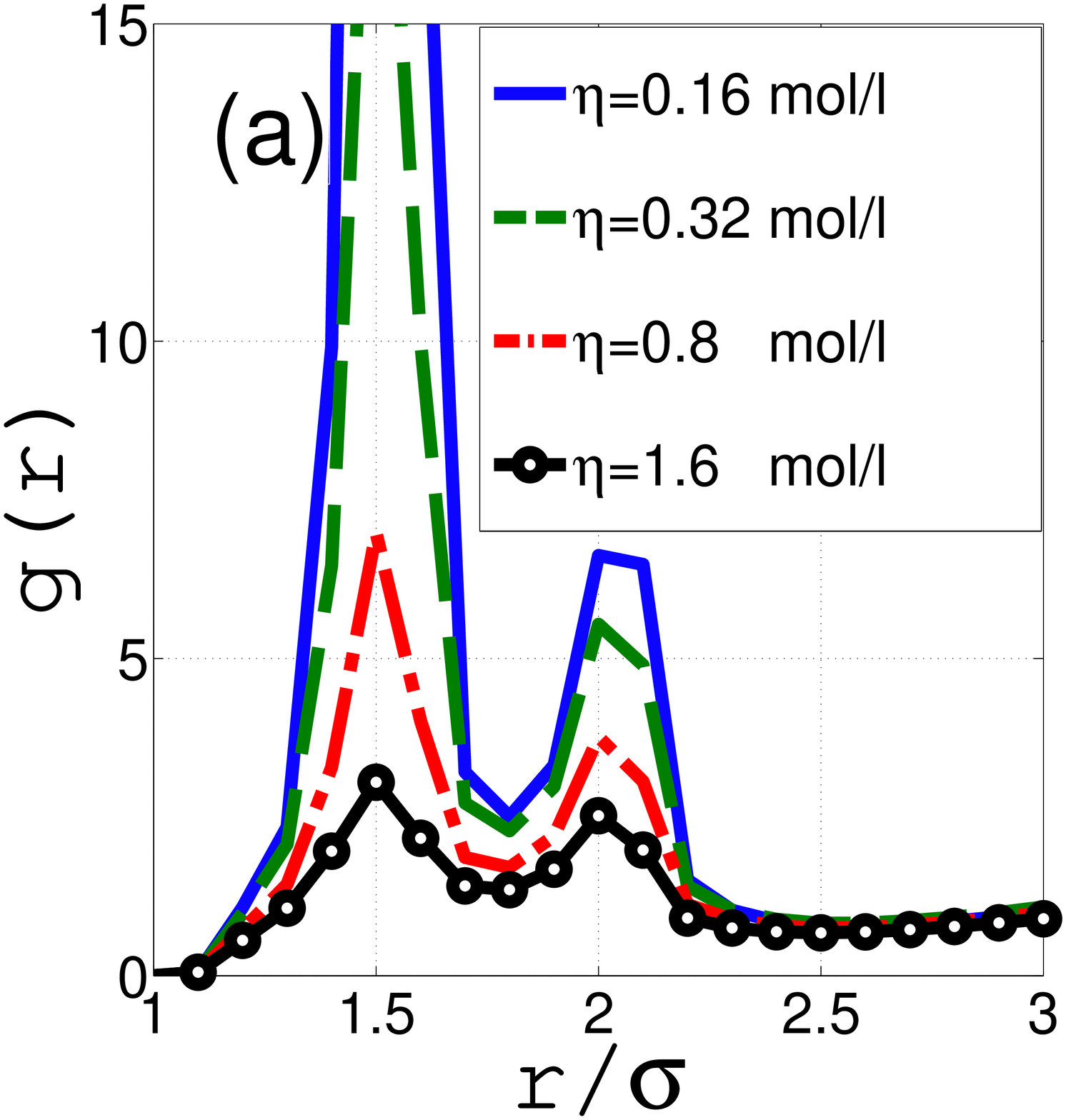}\\
\includegraphics*[width=0.65\textwidth]{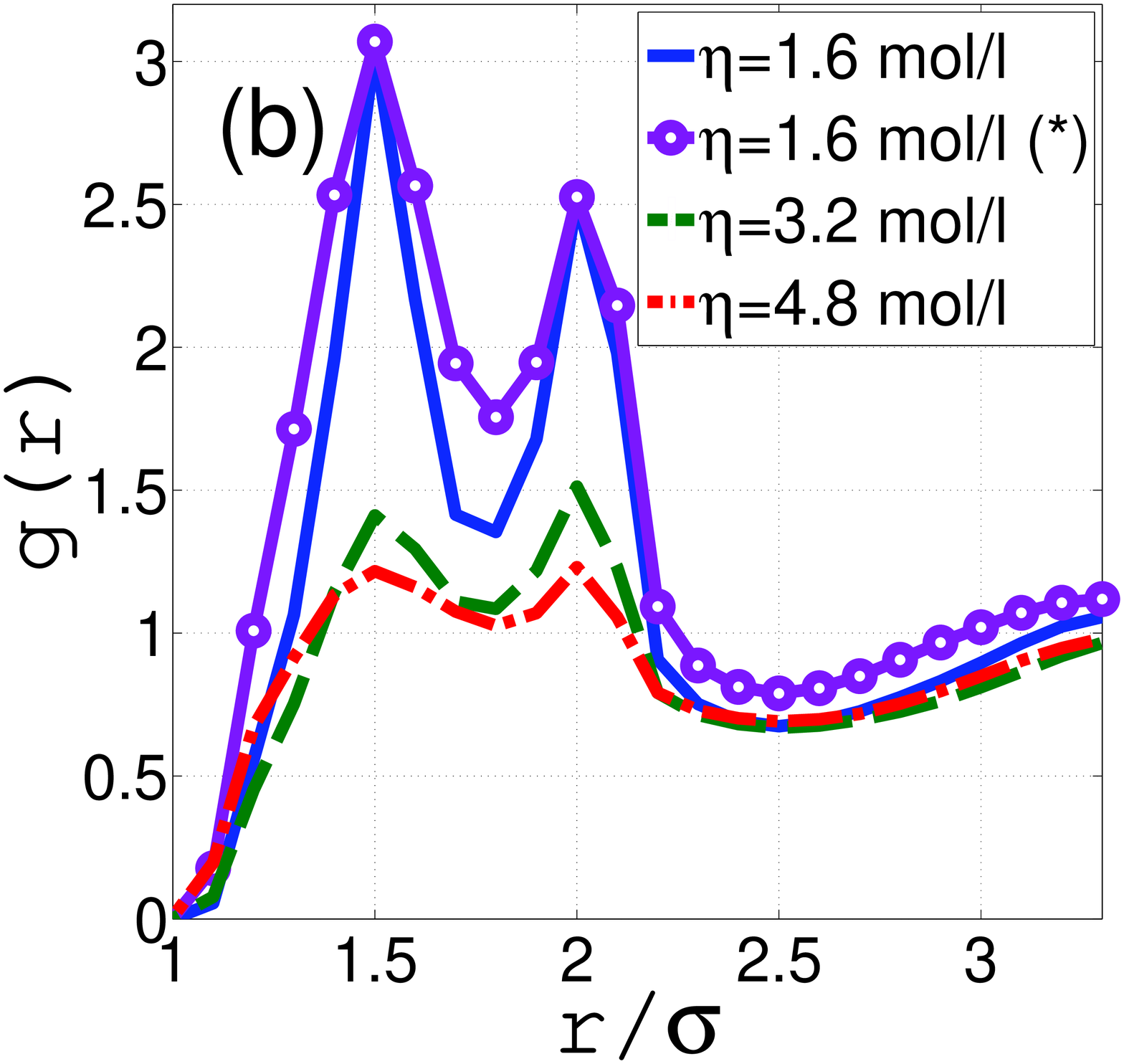}

\end{figure}

In the previous section, in which we varied the strength of the Coulomb interactions, we observed that
in the systems with the higher sulfonate concentration, the Coulomb correlations were suppressed. This suggested
that a study of the morphology of aggregates over a wider range of
concentrations would be helpful to understand the physics behind sulfonate
cluster formation. Simulation results for $g(r)$ for systems with
sulfonate concentrations $\eta$ varying between 1.6 and 2.5 mol/l are
shown in Fig.~\ref{fig9} for Runs 8 and 10-12.    Results for the much wider range of  0.16 to 4.8 mol/l are shown  in Fig.~\ref{fig9a}, but for these it was necessary to delete the backbone segments entirely in order to accommodate the very large concentration of side chains at the upper part of the range
constituting Run~13.   

The dipole-dipole correlations are seen to depend strongly on the molar
concentration of acidic groups. For a typical dilute case, Run~10 in
Fig.~\ref{fig9}, the first peak of $g(r)$ reaches a value $g(r=1.5\sigma)=37$,
indicating a strong condensation of nearly all the side chains. A
possible explanation for this effect can be drawn from the clustering
features of pure dipolar liquids. In a dilute system, where the average
interchain distance is larger than the gyration radius $R_{g}$ of
the side chain, the LJ interaction between the neighboring chains
is negligible. The main contribution to the energy in this case comes
from the electrostatic dipole-dipole interaction of head groups. Although
this interaction decays with a $1/r^{3}$ distance dependence, it
is not weak over distances comparable with the simulation box length
$L$, and is capable of forcing the chains to aggregate. However,
for a dense system, where the LJ contribution to the energy becomes
comparable to the dipolar term, the side chains start to lose their
rotational entropy, and thus cannot reorient themselves in order to
minimize the electrostatic energy of the aggregate. While the heights
of the maxima of $g(r)$ gradually decrease as the system becomes
more tightly packed, their positions do not change and remain fixed
at $r=1.5 \sigma,\,2 \sigma,$ and$\,3.4 \sigma$.

\begin{figure}

\caption{(Color online) Typical snapshots of system for Run~13 containing $\eta=0.4$mol/l
(case (a)) and $\eta=3.2$ mol/l (case (b)) head groups.
\label{fig15}}
\includegraphics*[width=0.65\textwidth]{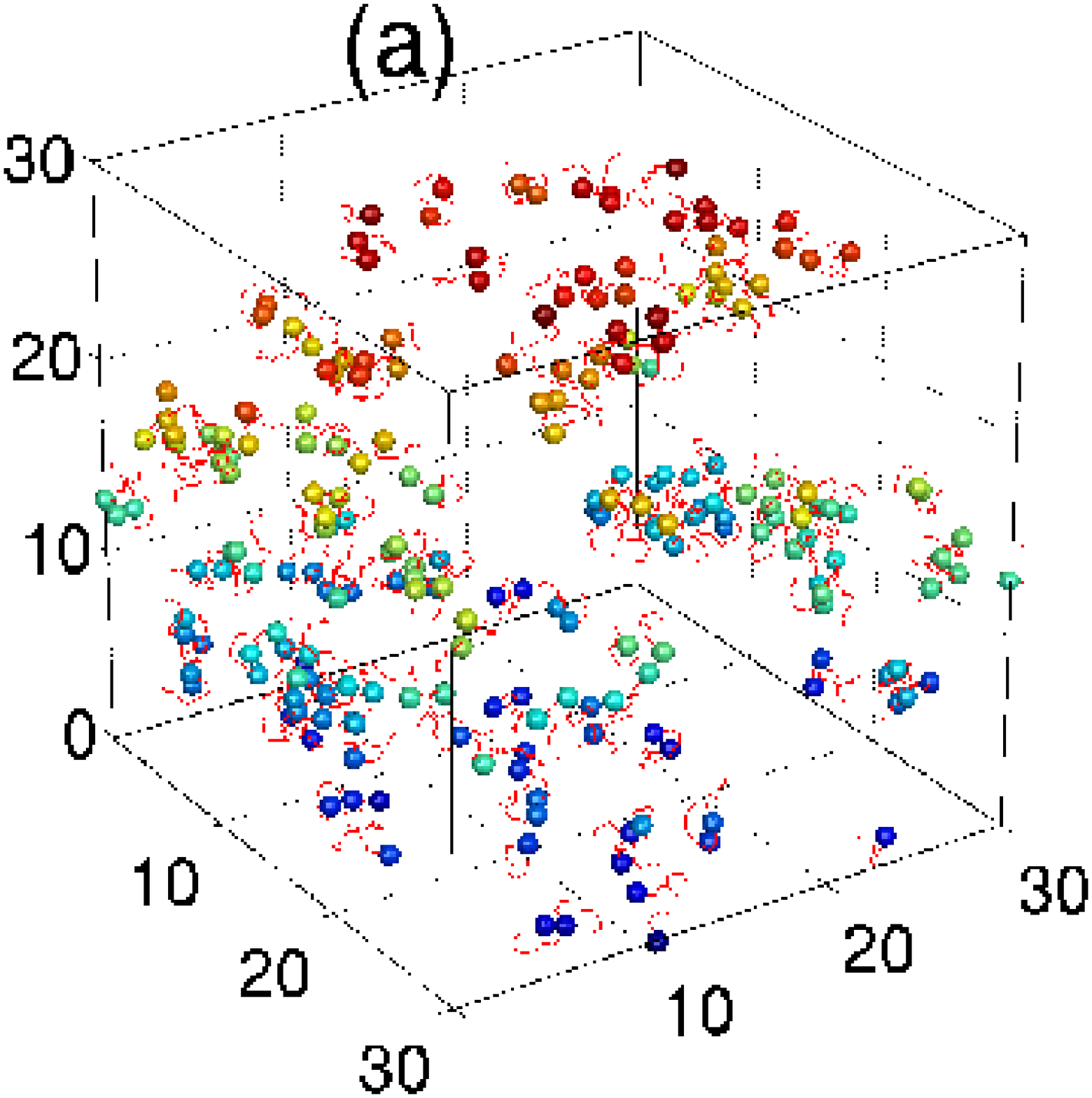}
\\ \includegraphics*[width=0.65\textwidth]{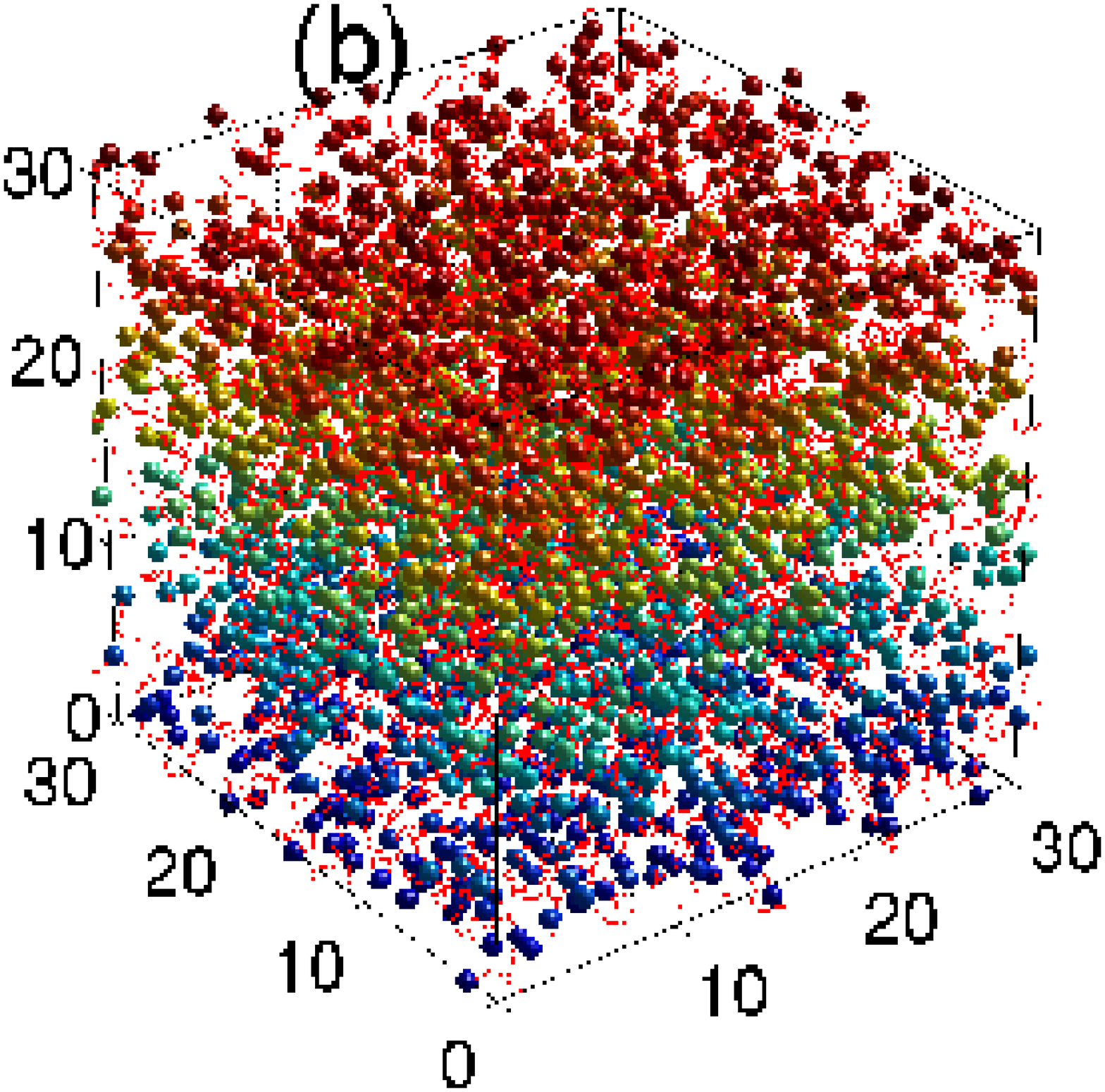}

\end{figure}

\begin{figure}

\caption{(Color online) A short-time average of 3D density of ionomer monomers for Run~13 and
  $\eta=2.4$ mol/l. The areas of low density are cut away.
\label{fig8}}
\includegraphics*[width=0.9\textwidth]{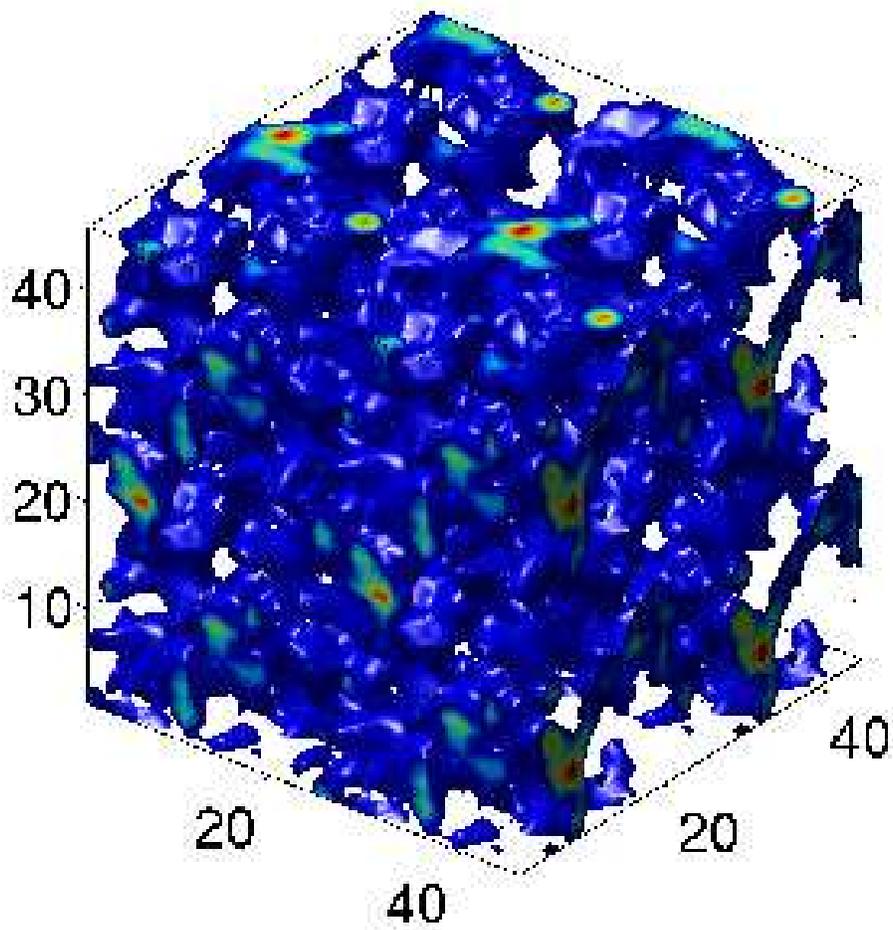}
\end{figure}

The line with circles in Fig.~\ref{fig9a} corresponds to the case where the concentration of backbone segments was double the case for Run~8, and corresponds to Run~9. It
is seen that the proton pair distribution function is hardly affected
by the existence of the extra hydrophobic segments. However this conclusion
is not necessarily true for a system with the more attractive LJ parameters
for HH and HP interactions that will be considered in section D. Both the snapshot
picture in Fig.~\ref{fig15} and the 3D monomer density in Fig.~\ref{fig8}
show microphase separation.

\begin{figure}

\caption{(Color online) Structure factor $S(q)$ of the head groups for Run~3
versus wave number $q\sigma$. \label{figsk}}
\includegraphics*[width=0.9\textwidth]{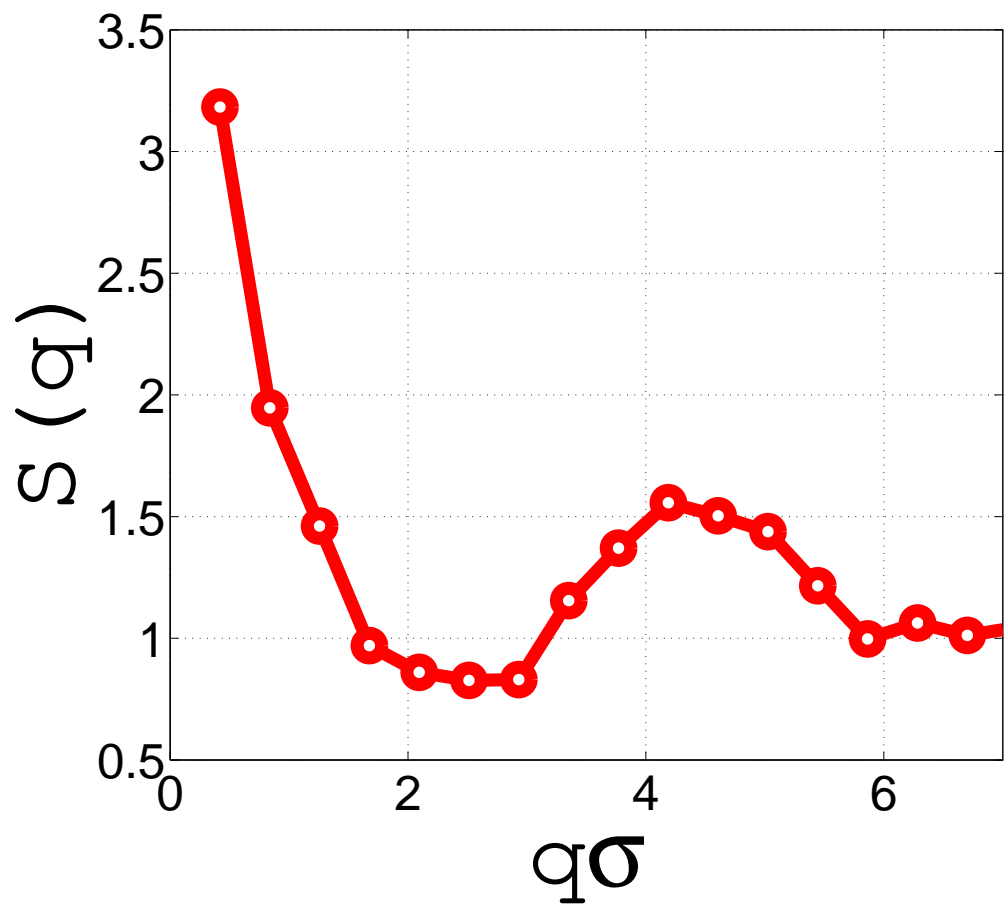}

\end{figure}

Another way to examine the head group clustering is to calculate the
partial structure factors $S(q_{j})$ ($j=x$, $y$, $z$) of the
protons. Because these did not exhibit any marked anisotropy, an averaged
$S(q)$ is plotted in Fig.~\ref{figsk}. The position of the peak
at $q\sigma=4.2$ is associated with a correlation distance, $d=1.5\sigma$
in the sample structure via the Bragg relationship $d=2\pi/q$. A
sharp growth at small $q$ of the values of $S(q)$ in Fig.~\ref{figsk}
results from microscopic aggregation of hydrophilic and hydrophobic
monomers in the system.

To summarize the tendencies of dipole-dipole correlations presented
in this and the previous section, we conclude that the hydrophilic head group aggregation
takes place preferentially at low dielectric permittivity $\epsilon$
and low densities of sulfonate groups.

\subsection{Role of side chain size $N_{M}$.}

Although the bulk of this work describes the conformational structures
of side chains of length $N_{m}$=10, it is worth while investigating
systems with shorter side chains, since recent experiments \cite{arcella}
show a high proton transfer rate in ionomer membranes with shorter pendant
segments.

\begin{figure}

\caption{(Color online) Proton-proton pair distribution function $g(r)$ for
a system of side chains for Run~14. The number of monomers per side
chain segment $N_{m}$ is changed from 2 to 10. \label{fig12}}
\includegraphics*[width=0.9\textwidth]{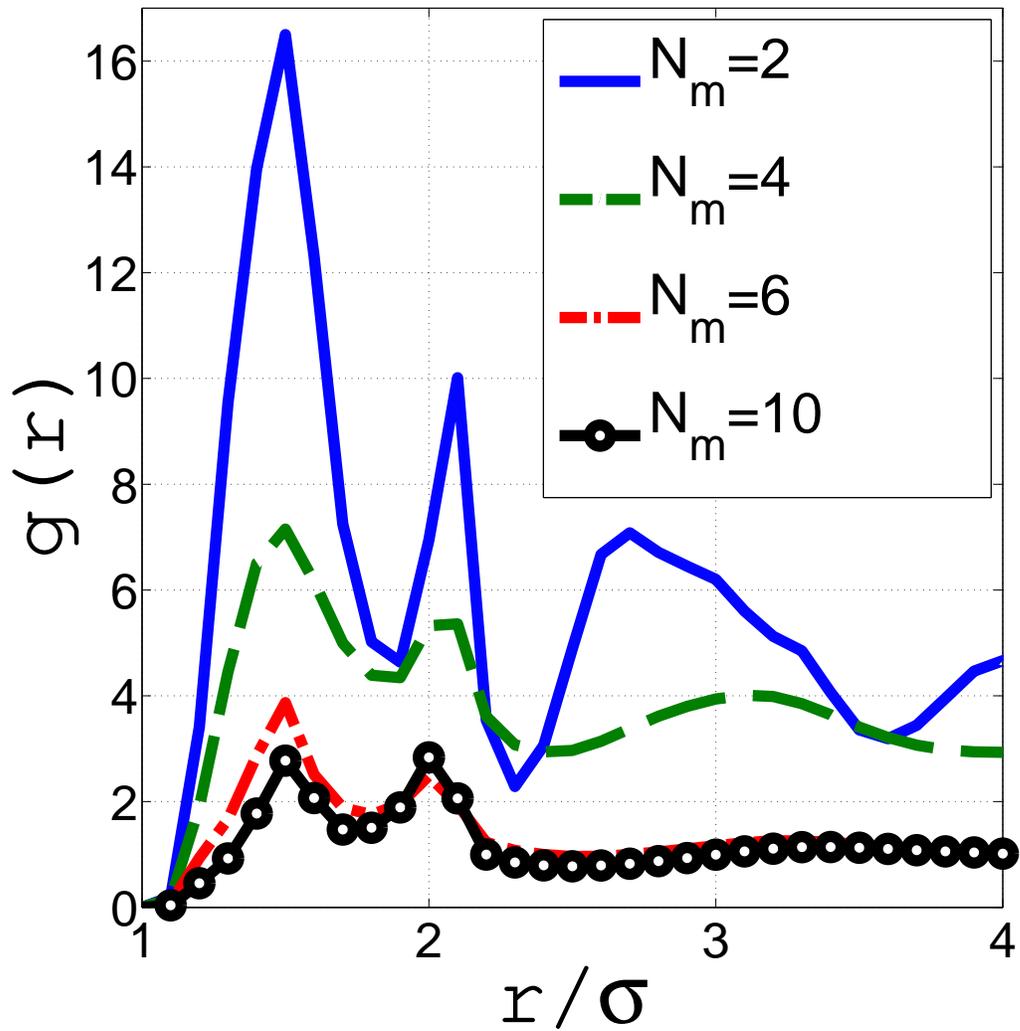}

\end{figure}

This section contains simulation results for different side chain
lengths, $N_{m}=N_{P}+N_{H}$. The length of the hydrophilic part (the
number of hydrophilic monomers) is fixed at $N_{P}=2$ and the length
of the hydrophobic tail (the number of hydrophobic monomers) $N_{H}$
is varied between 0 and 8 (which corresponds to the variation of total
side chain length $N_{m}$ between 2 and 10) in order to reveal how
the side chains' hydrophobic tail affects the pair distribution function
of protons. No backbone material was included in these systems.  Simulation results for Run~14 are shown in Fig.~\ref{fig12}.
Note that the case $N_{m}=2$ corresponds to a system of dipolar dumbbells.
Even a short hydrophobic tail with $N_{m}=4$ strongly suppresses
the aggregation of head groups. Thus the hydrophobicity of side chains,
or the length of hydrophobic tail, is a strong obstacle to head group
clustering. In other words, the dipole-dipole attraction between sulfonates
of longer side chains is not sufficient to produce a nanophase separation.
This conclusion is in contrast to the suggestion of Eisenberg \cite{eisenberg}
that each ionic group in the vicinity of a cluster of sulfonate groups
is drawn in by dipolar attraction to form a depleted zone free of
charge.

\subsection{Ionomer morphology for strong hydrophilic and hydrophobic interactions. }

\begin{figure}

\caption{(Color online) Proton-proton pair distribution function $g(r)$ for
Run~1 (solid line), Run~2 (dashed line) and Run~3 (dot-dashed
line). Thick lines: $\varepsilon_{LJ}=9k_{B}T$ for
hydrophobic-hydrophobic (HH) interaction, $\varepsilon_{LJ}=3k_{B}T$
for hydrophobic-hydrophilic 
(HP) and hydrophilic-hydrophilic (PP) interactions. Thin lines:
$\varepsilon_{LJ}=3k_{B}T$ for HH, $\varepsilon_{LJ}=1k_{B}T$ for
HP and PP interactions. \label{fig16}}
\includegraphics*[width=0.8\textwidth]{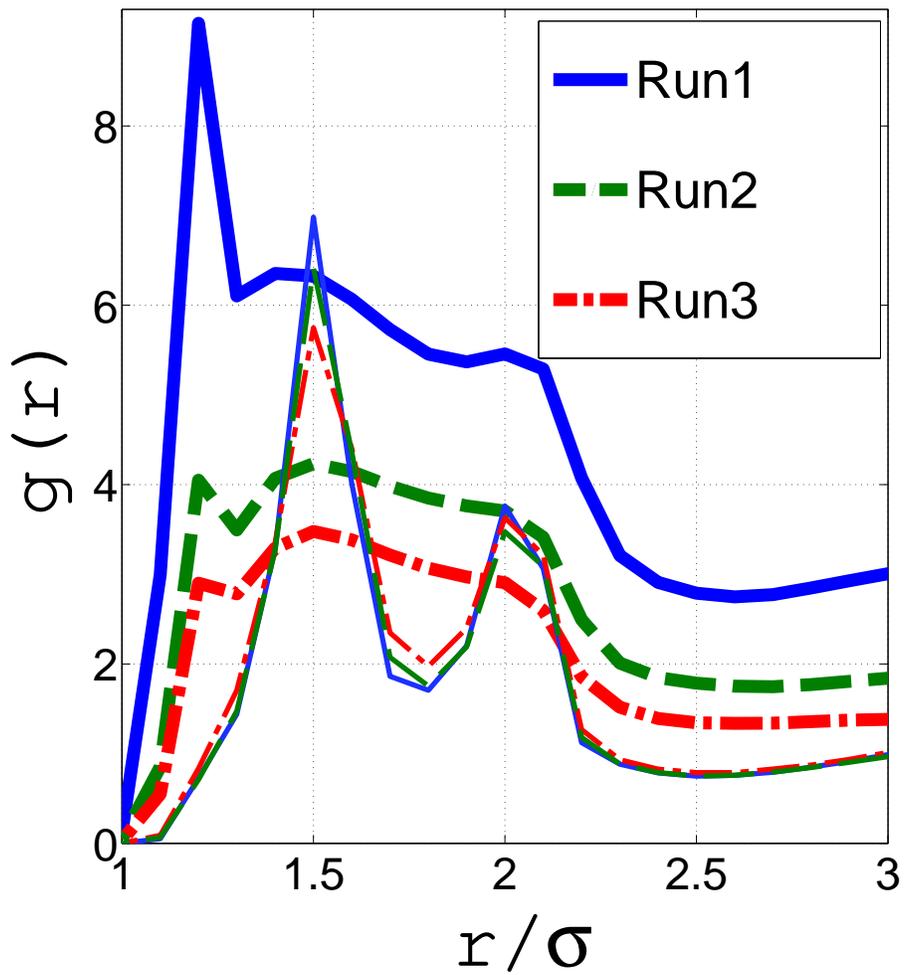}

\end{figure}

\begin{figure}[h]

\caption{(Color online) Snapshot pictures of systems for Run~3 and $n_b\psi=$2.
Case (a): $\varepsilon_{LJ}=3k_{B}T$ for HH, $\varepsilon_{LJ}=1k_{B}T$
for HP and PP interactions. Case (b): $\varepsilon_{LJ}=9k_{B}T$
for HH and $\varepsilon_{LJ}=3k_{B}T$ for HP and PP interactions.
In the case (b) the sulfonate groups are mostly on the surface
of clusters.}
\label{fig18}
\includegraphics*[width=0.6\textwidth]{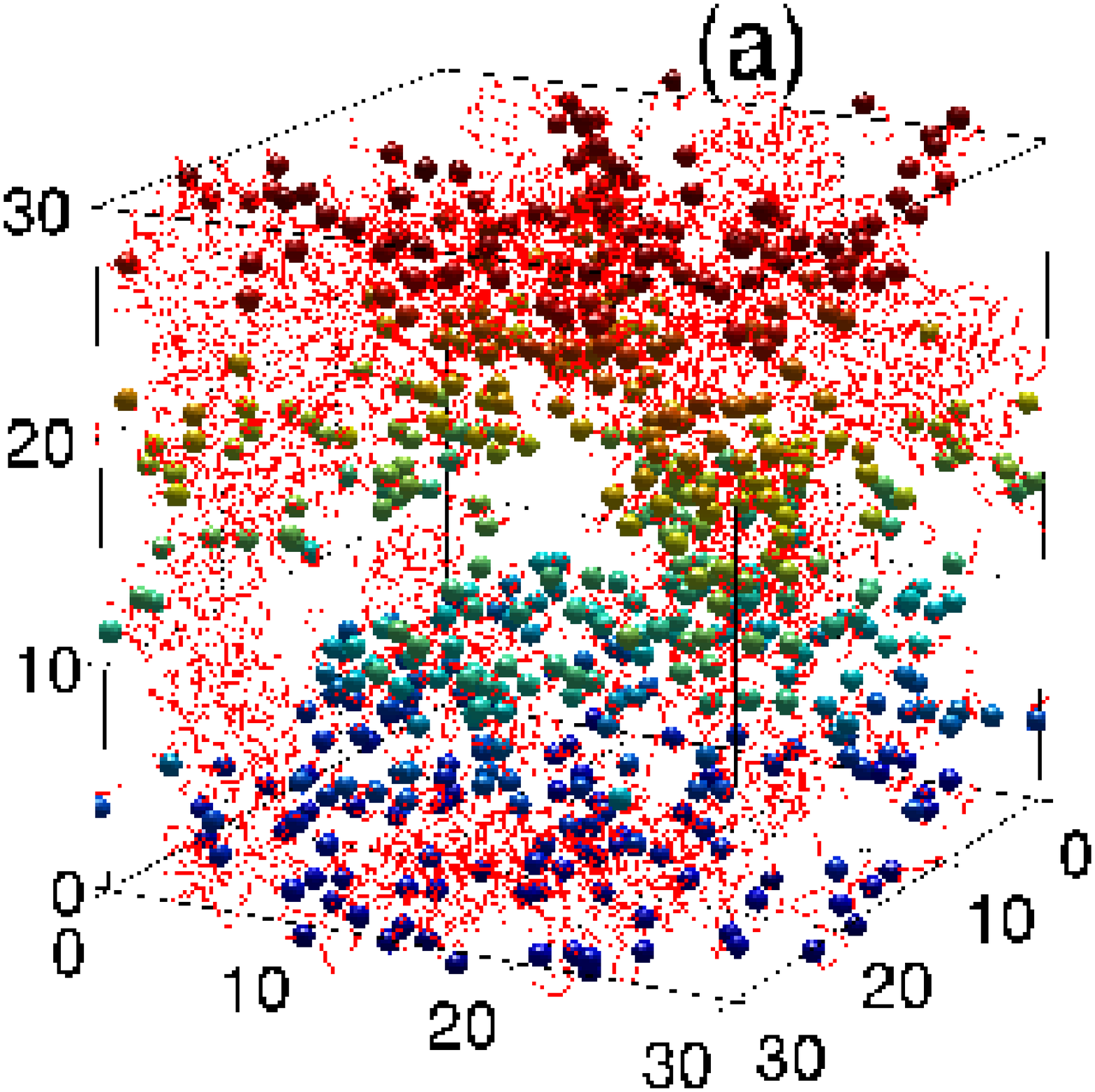}
\\ \includegraphics*[width=0.6\textwidth]{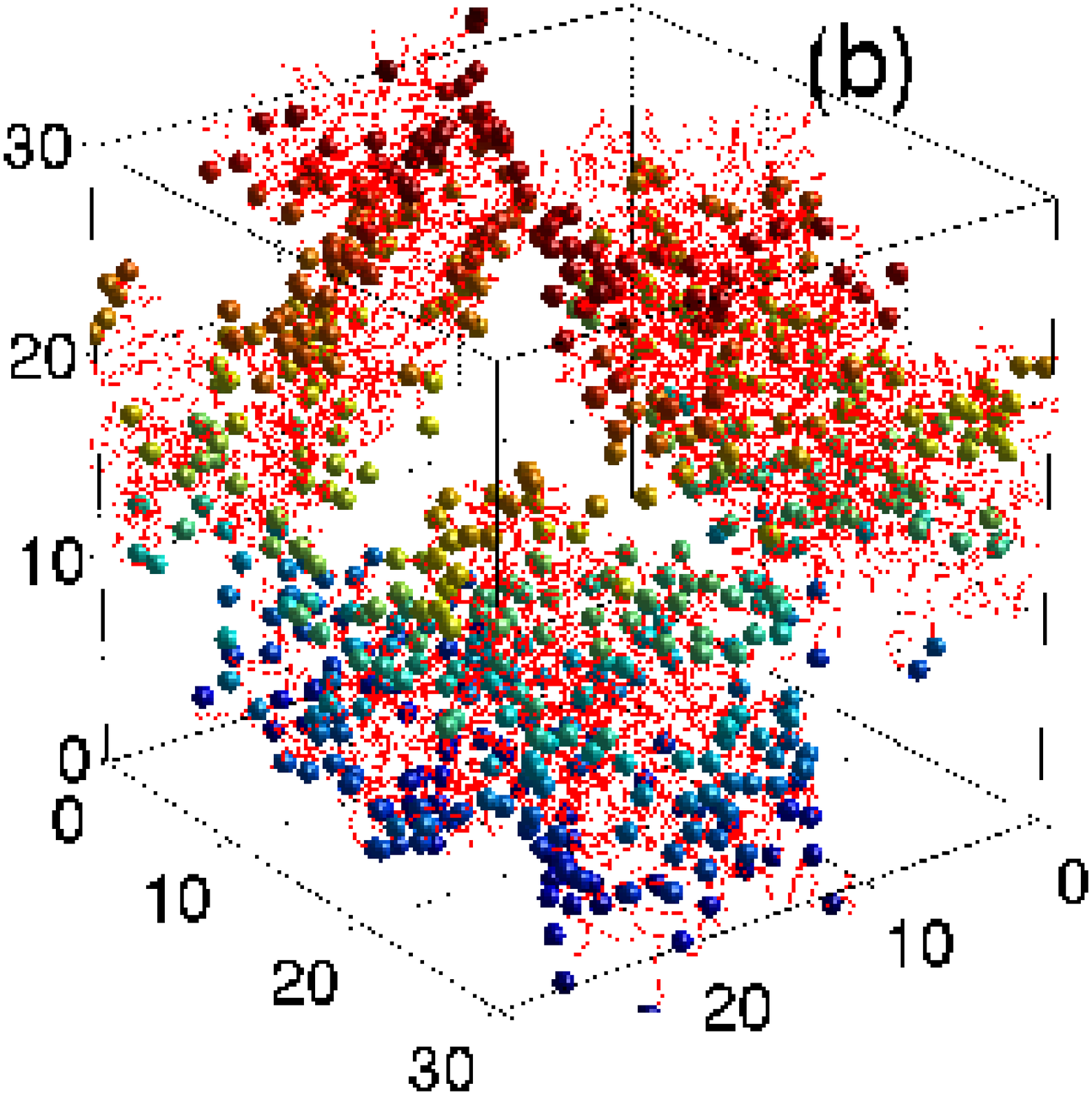}
 
\end{figure}

One intriguing question in fuel cell technology is how the morphology
of nanophase separation in PEM membranes depends on the polymeric
structure of Nafion. In other words, what is the optimal comb-like
structure for the ionomer to be a good material for a fuel cell? Using
our segmental model for Nafion we can mimic different comb-like Nafion
molecular conformations by changing the parameter $\psi$, which denotes the ratio of the number of sulfonate groups to the number of backbone monomers.
The impact of $\psi$ on the distribution function $g(r)$ was shown
to be negligible for $\varepsilon_{LJ}=3k_{B}T$ for HH and $\varepsilon_{LJ}=1k_{B}T$
for both HP and PP interactions in Fig.~\ref{fig6} for Runs 1-6. In
order to explore more generally the effect of changing $\psi$, we
exaggerate the hydrophilic and hydrophobic interactions by tripling
the interaction constants  $\varepsilon_{LJ}$.  The Results for Runs
1-3 presented in Fig.~\ref{fig16} for $\varepsilon_{LJ}=9k_{B}T$ 
for HH, and $\varepsilon_{LJ}=3k_{B}T$ both for HP and PP interactions,
now show that the addition of backbone chain segments
brings more conformational disorder into the head group correlations.
This is also evident from the system snapshots for Run~3 in Fig.~\ref{fig18}
for the two different strengths of Lennard-Jones interaction parameters. In other words,
a strong interaction between hydrophobic and hydrophilic monomers
can disrupt the side chain aggregation, causing all the side chains to be
evenly distributed on the surface of backbone aggregations, as in Fig.~\ref{fig18}(b).
This distribution will have a globular, lamellar, or cylindrical-rod form.

\subsection{End group attraction between side chains}

\begin{figure}[h]

\caption{(Color online) Proton-proton structure factor $S(q)$ for Run~8. Dipole
strength $D_{r}$ is scaled down from 1 to 0. \label{fig1}}
 \includegraphics*[width=0.75\textwidth]{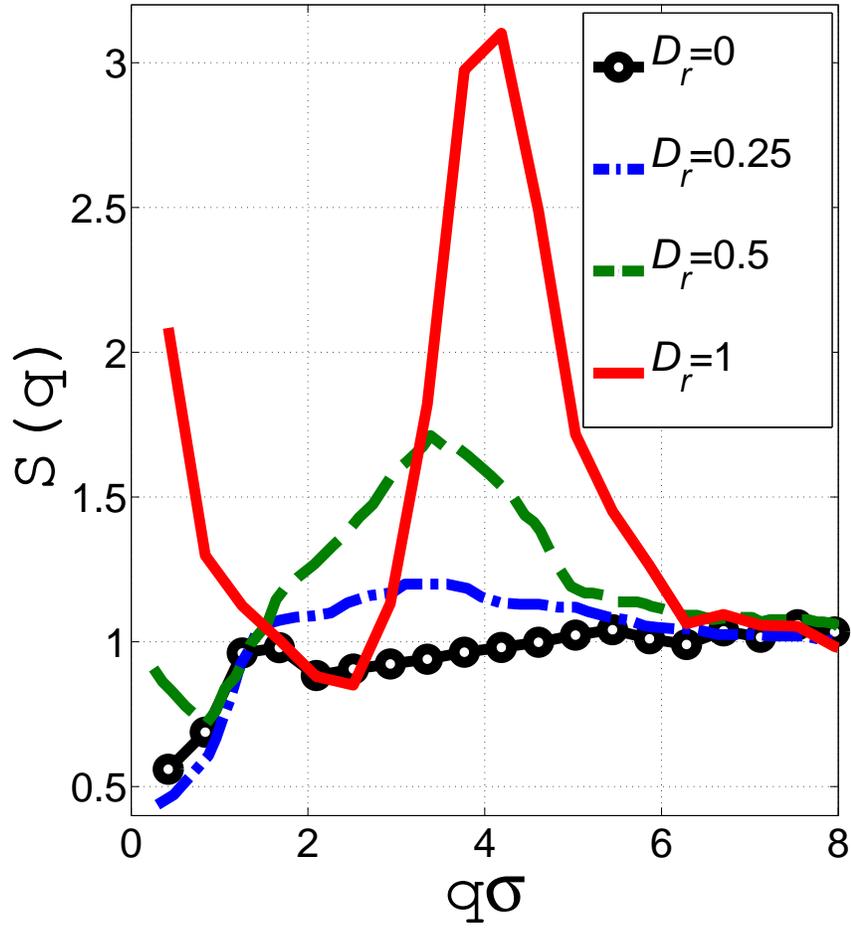}
\end{figure}

To complete this study of the effects of interactions on morphology,
we look at the effect of both decreasing and increasing the
interaction between head groups.  We have already seen in section IIIA
some of the consequences of reducing the Coulomb interaction by
increasing the dielectric constant $\epsilon$.   In the absence of
interactions other than the dipole-dipole interaction, this is
equivalent to a reduction in the strength of the dipole moments.   In
Fig.~\ref{fig1} we show the effect on the  structure factor 
$S(q)$ of the protons of a reduction in the strength of the dipole
moments by various factors $D_r$ (Run~8). When 
the parameter $D_r$ is small, there is no phase separation in the ionomers,
and the function $S(q)$ has a liquid-like structure with a single
maximum. The behavior of $S(q)$ is changed for an unreduced dipole
strength $D_r$=1 in a manner similar to the behaviour of $S(q)$ in the less dense system shown in
Fig.~\ref{figsk}: there is a strong peak at $q\sigma$=4.2, and
a sharp growth at small $q\sigma$. The latter, again, is an indication
of a strong microscopic separation of hydrophilic and hydrophobic
monomers in the system.

It is clear that the effective attraction between the acidic groups
of side chains cannot be accounted for by the sulfonate-hydrogen
dipole moment alone. The sulfonate end group has its own internal
dipole moment due to the charge distribution between the sulfur atom
and the three oxygen atoms bound to it. Partial charges on the other side
chain monomers can also contribute to the effective attraction between
head groups. To take all these contributions into account at a phenomenological
level we introduce an artificial attractive interaction between the
end groups of the side chains, in an approach similar to that employed
by Shirvanyanz \cite{shirvanyanz}. For this purpose we use a Yukawa-type
attractive potential, \[
F(r)=-(Z^{2}e^{2}/\epsilon r)\exp\left(-r/r_{D}\right)\,\,\,{\textrm{for}}\,\, r>\sigma\,;\,\,\,\,\, F(r)=\infty\,\,\,\,\,{\textrm{for}}\,\,\, r<\sigma.\]
 Here $Z$ is an effective Yukawa charge for the head group sulfonates.
For the sake of simplicity, we restrict ourselves to the case of constant
Debye screening length $r_{D}=4\sigma$. In fact, the parameters $Z$
and $r_{D}$ determine the stability, size and shape of the aggregates
forming in the system. The simple attractive Yukawa term can serve
as a convenient basis for interpreting all the important features
that can be observed in real systems or predicted by statistical mechanical
theories.

\begin{figure}[h]

\caption{(Color online) Proton-proton pair distribution function $g(r)$ for
Runs 7-9 and $Z=0.1$ (a) and $Z=0.5$ (b).
$Z$ is a fictitious charge on the head groups which causes an artificial
Yukawa like attraction between sulfonates. Thick (thin) lines are for
$D_r=0$ ($D_r=0.5$). Note that the scales on $y$-axis for short-range and
long-range behavior of $g(r)$ (shown as an inset) are different. For
other details see the 
text. \label{fig19}}
\includegraphics*[width=0.7\textwidth]{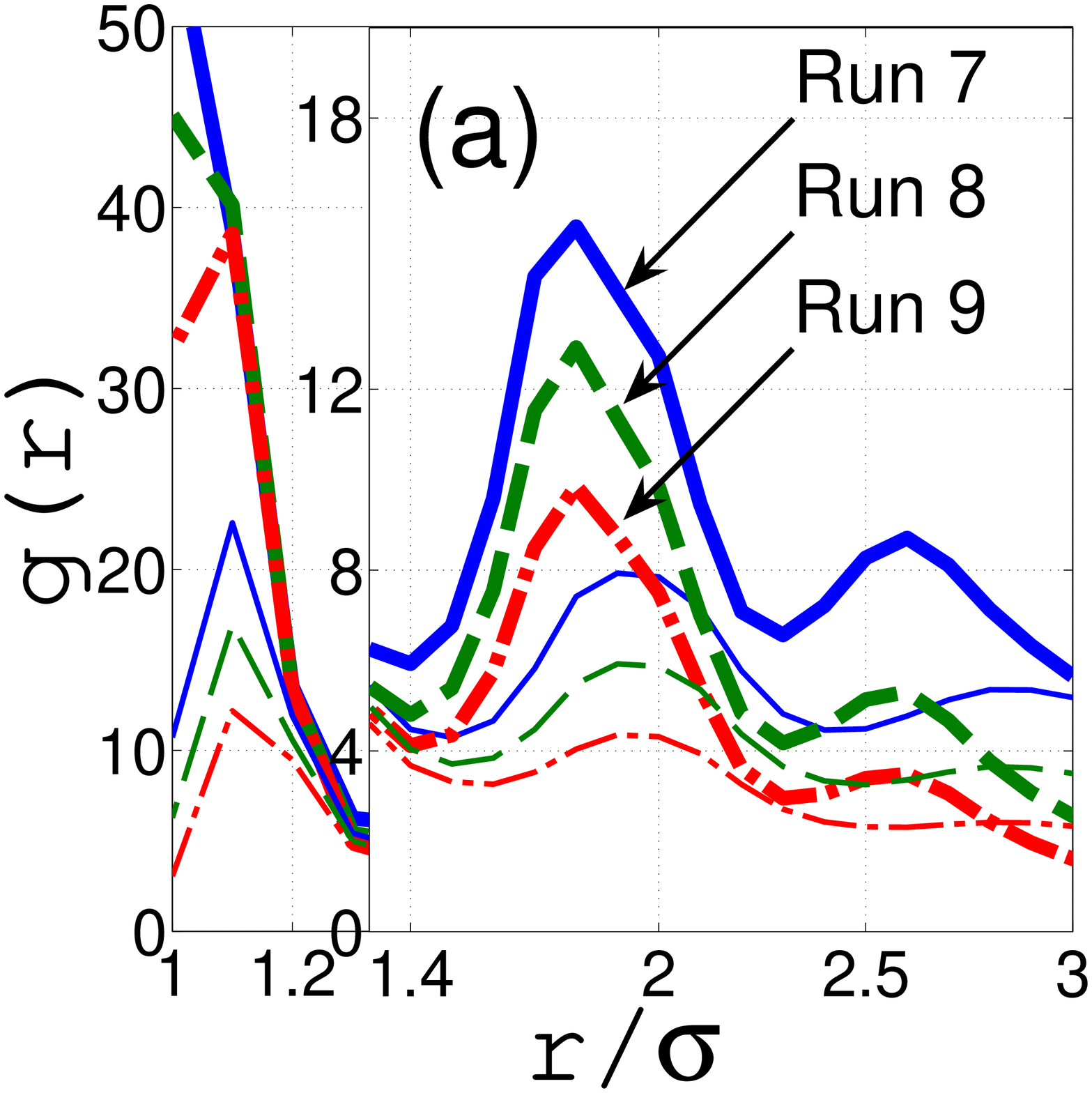}
\\ \includegraphics*[width=0.7\textwidth]{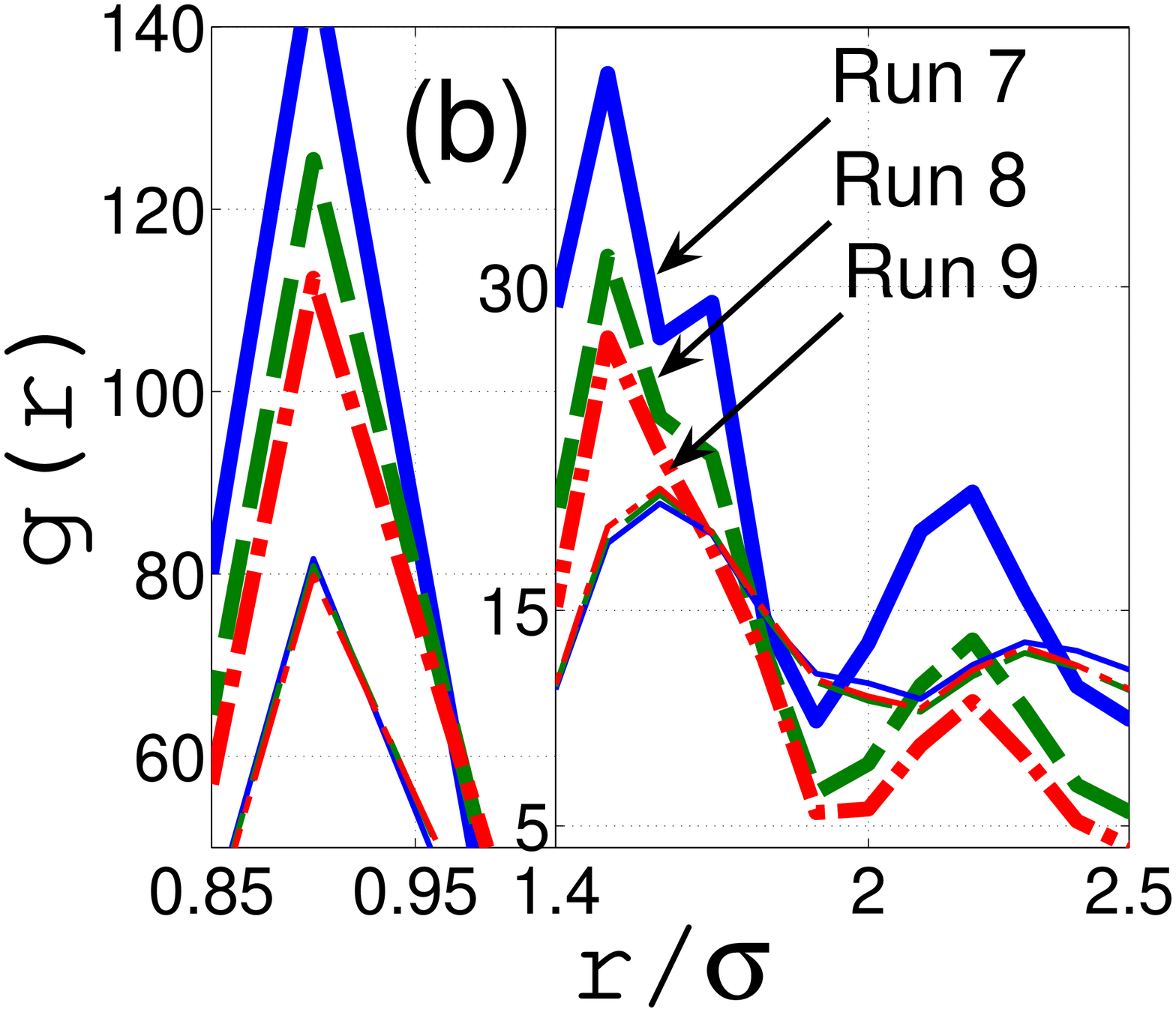}

\end{figure}

\begin{figure}[h]

\caption{(Color online) Snapshot pictures of systems for Run~9, $n_b\psi=$2,
$D_r$=0, $Z$=0.1 (case (a)) and $Z$=0.5 (case (b)). \label{fig23}}
\includegraphics*[width=0.7\textwidth]{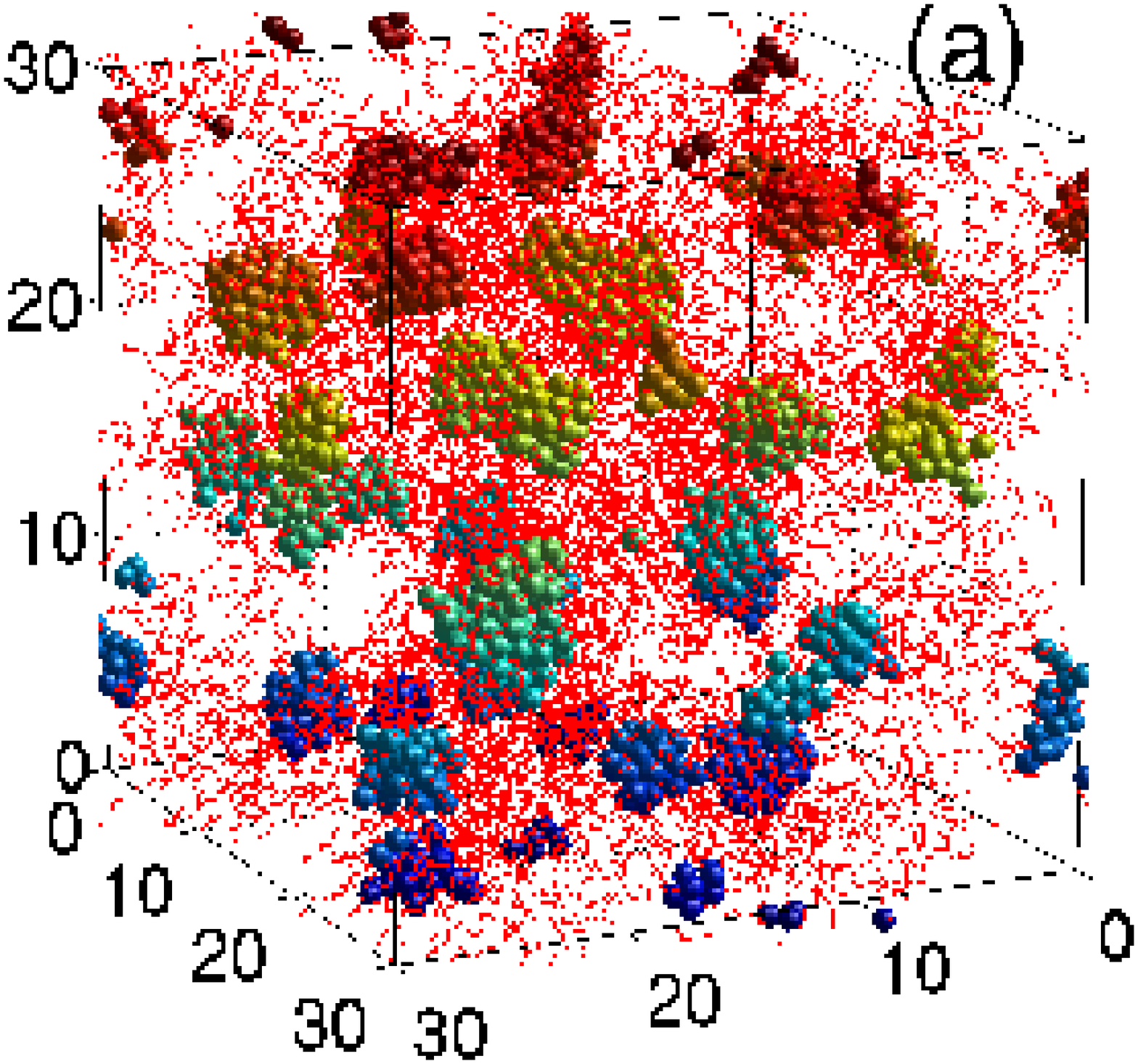}
\\ \includegraphics*[width=0.7\textwidth]{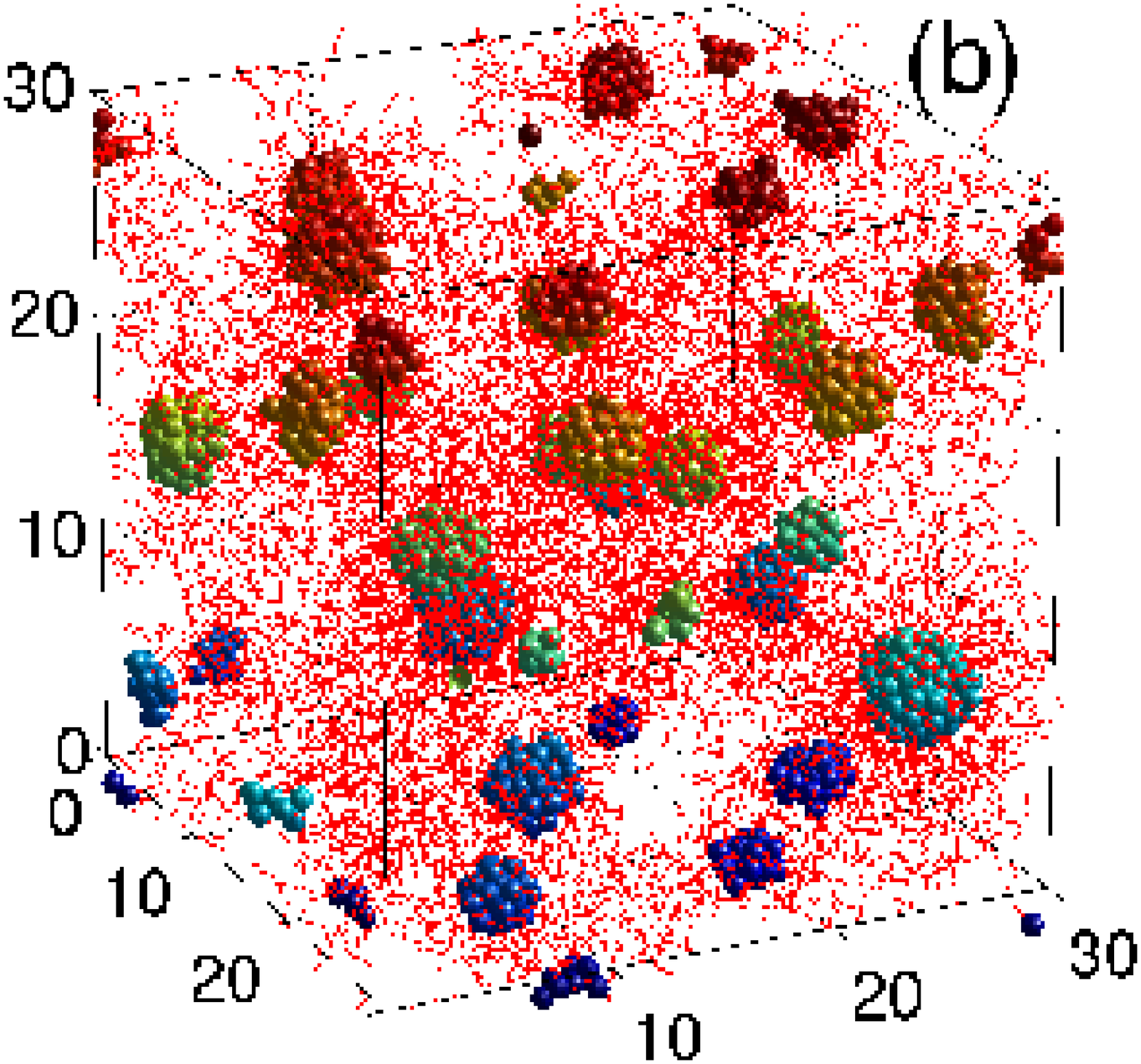}

\end{figure}

In the absence of electrostatic interactions in the system, i.e. when
the head group dipole moment vanishes and $D_r=0$, the ionomer system becomes
spatially inhomogeneous. The pair distribution functions $g(r)$ for
Runs 7-9, plotted in Fig.~\ref{fig19} for $Z=0.1$ (case (a)) and
$Z=0.5$ (case (b)) show strong structuring. The aggregation is substantially
stronger for a system missing backbone polymer (Run~7) than for a  system with a full complement of backbone monomers (Run~9),
as seen from a comparison of the respective lines in Fig.~\ref{fig19}.
There is a clear tendency to micellar arrangement, which is characterized
by some short-range order. Snapshot pictures, given in Fig.~\ref{fig23},
show formation of a space-filling web-like network of cluster-like
aggregates surrounded by neutral polymer sections. The size of these
clusters grows as the density of side chains increases. A comparison
of Figs.~\ref{fig23}(a) and \ref{fig23}(b) reveals that the effective
Yukawa charge $Z$, or in other words, the strength of the the screened
Coulomb attraction, has a strong impact on the shape of the clusters.
The structure of the system for a moderate Yukawa attraction  of $Z =
0.1$ between
the end groups is mobile: the aggregates can diffuse,
decay and reappear again. They can also create temporary bridge-like
connections. Therefore the nanophase separation of the head groups,
shown in Fig.~\ref{fig23}(a) can be viewed as an example for a Gierke-like
sulfonate aggregation. However, a strong attraction between the end
groups of the side chains, the case $Z=0.5$ in Fig.~\ref{fig23}(b),
results in a frozen network of globular clusters of diameter roughly
1.5 nm in the polymer matrix. This cluster picture resembles the ion
segregation phenomenon proposed by Eisenberg \cite{eisenberg}.

\begin{figure}[h]

\caption{(Color online)  Snapshot pictures of systems for Run~9, $n_b\psi=$2,
$D_r$=0.5, $Z$=0.1 (case (a)) and $Z$=0.5 (case (b)).  \label{fig24}}
\includegraphics*[width=0.7\textwidth]{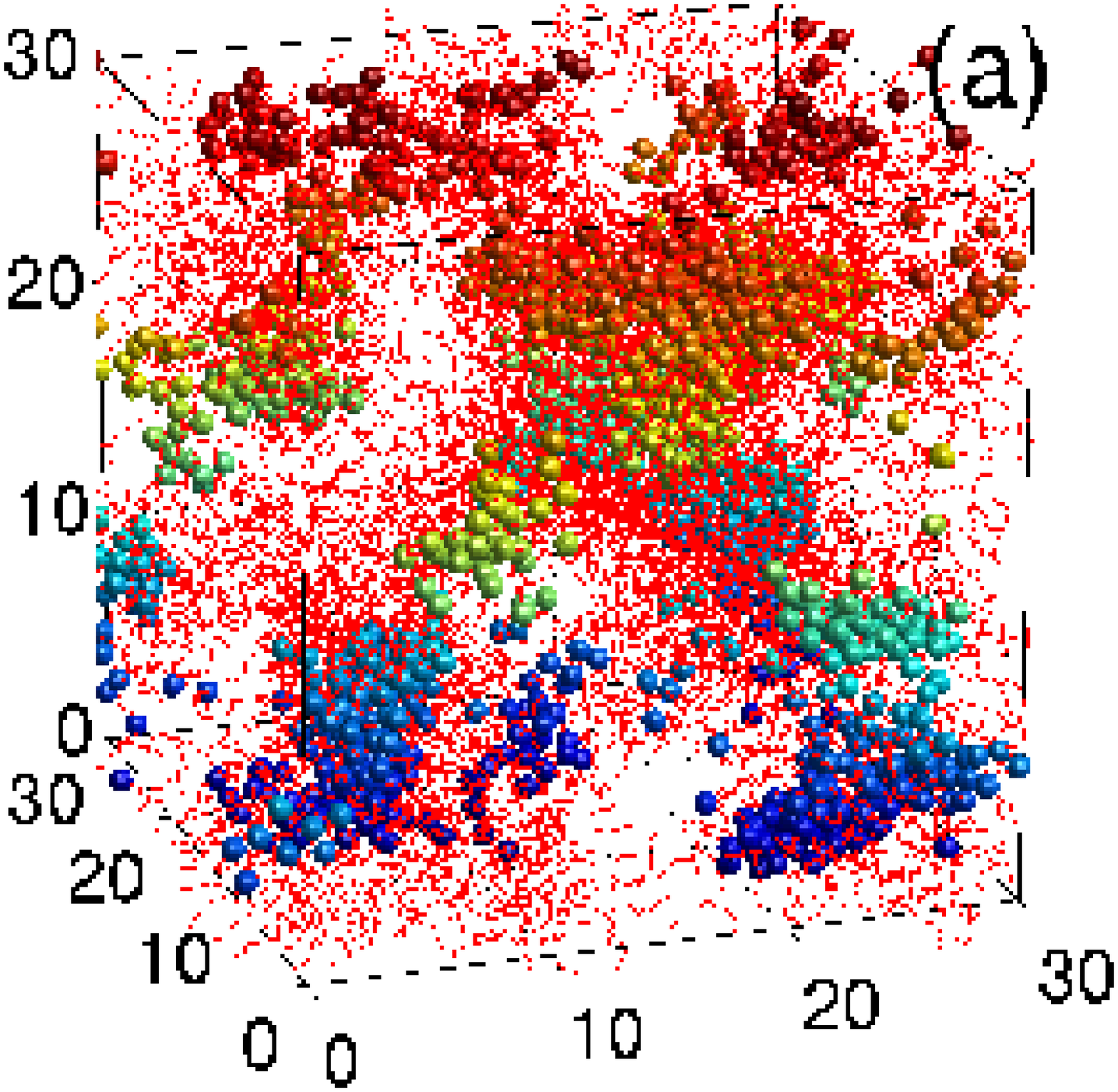}
\\ \includegraphics*[width=0.7\textwidth]{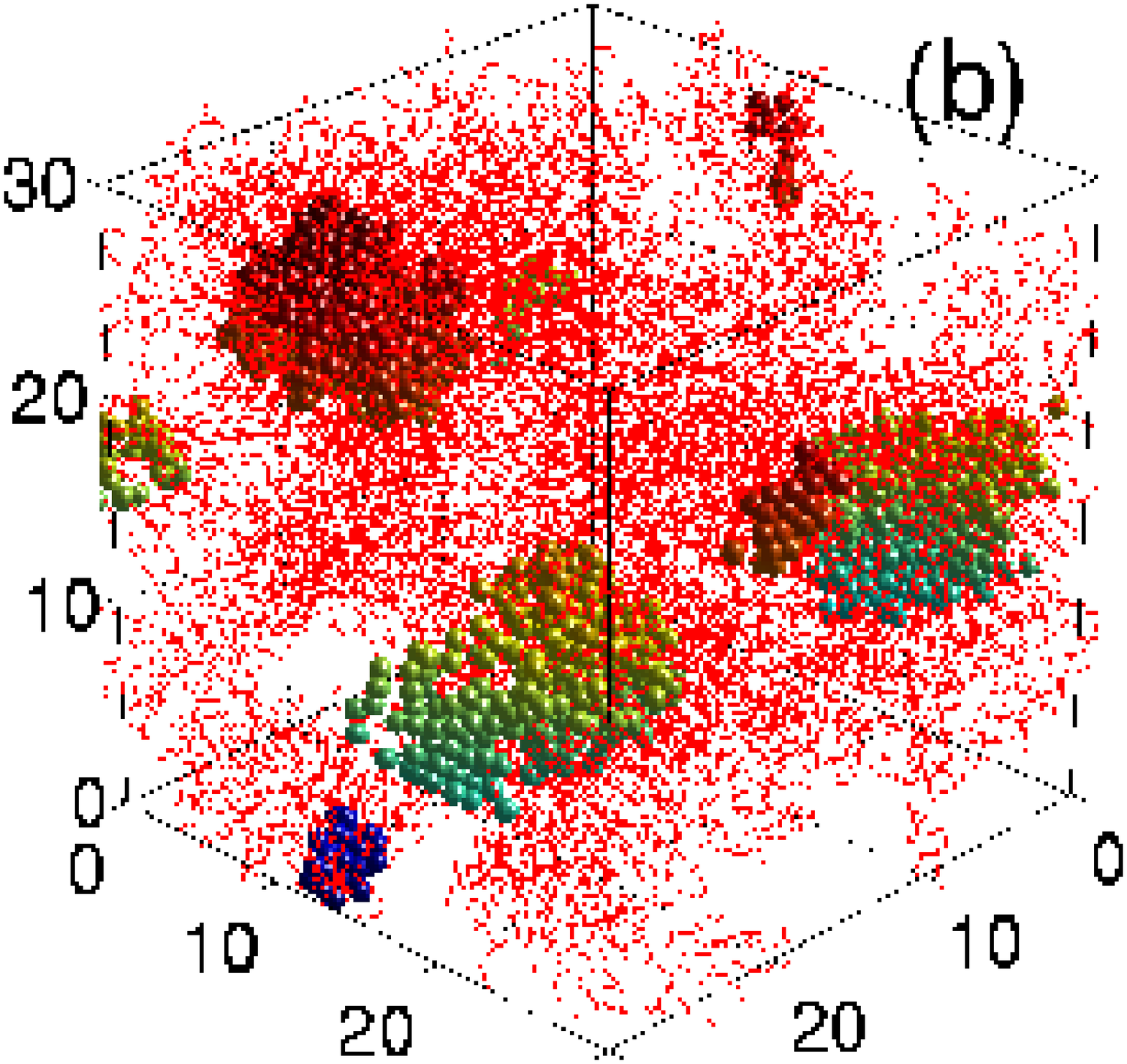}

\end{figure}

\begin{figure}[h]
\caption{(Color online) Snapshot of a system for Run~9, $n_b\psi=$2, D=0.5, $Z$=0.
\label{fig20}}
\includegraphics*[width=0.9\textwidth]{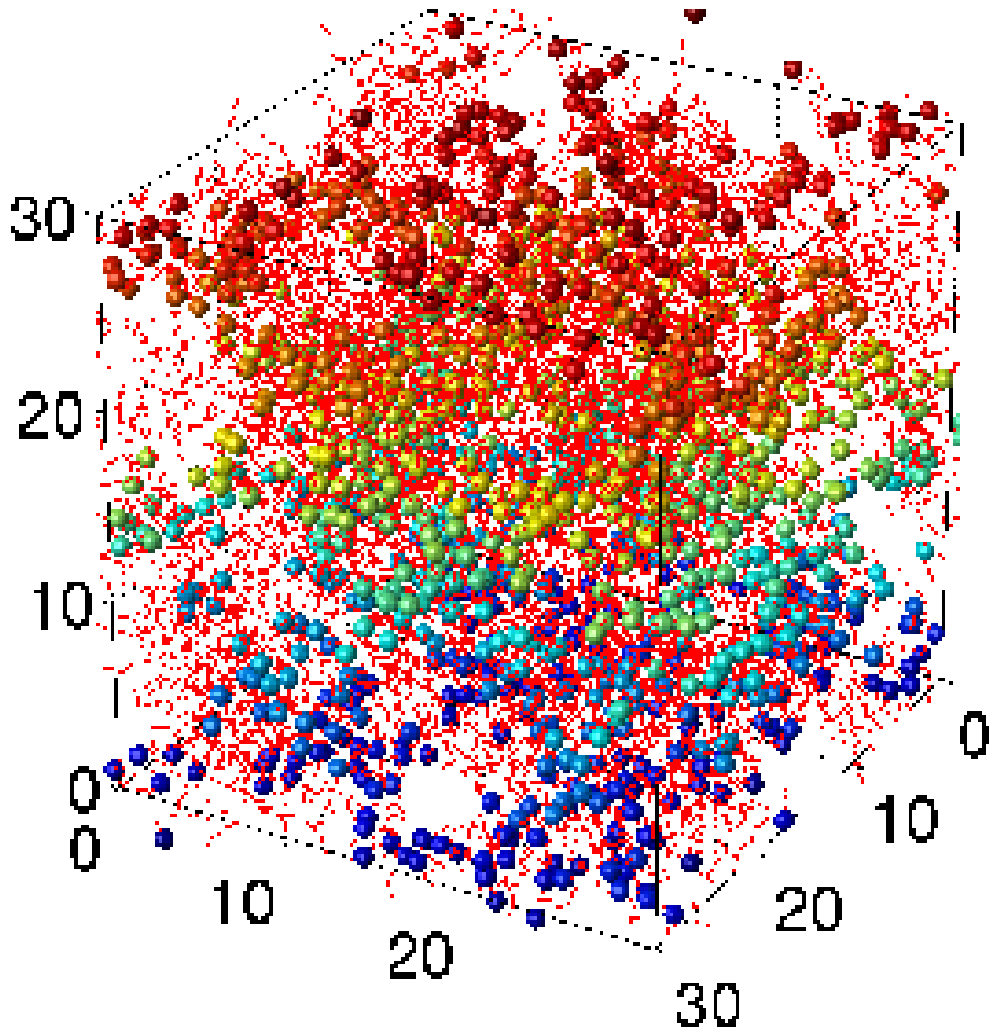}

\end{figure}

When the dipole moment of the sulfonate head groups is restored, in
this case with a value of $D_r=0.5$, the globular micelles form super-clusters
of diameter 3-4 nm , as seen is Fig.~\ref{fig24}(b) for $Z=0.5$.
On the other hand, this nonzero dipole moment facilitates the formation
of percolated cluster structures for $Z=0.1$, as in Fig.~\ref{fig24}(a).
The inclusion of the dipole moment of the sulfonate head groups thus
changes the conformational structure of clusters from a Gierke model
of spherical aggregations to a Gebel model of elongated aggregations.
A snapshot picture of the same system but with no artificial attraction
between the head groups is given in the Fig.~\ref{fig20}. It is
clear that the introduction of a screened Coulomb attraction results
in the formation of compact and interconnected clusters of sulfonate
head~groups.

\section{conclusions}

In this study we have explored the issue of nanophase separation in
dry Nafion$^{\tiny\textregistered}$ films by means of simulations
within a simple dipolar-head-group model for the side chains.  Our procedure involved detaching the side chains from the backbone and cutting the backbone into segments to speed equilibration, and then reassembling the macromolecule by means of a strong imposed attractive force between the cut ends of the backbone, and between the non-ionic ends of the side chains and the midpoints of the backbone segments.  

   As we
vary the interaction parameters and dielectric permittivity of the
medium we find different structures to be formed by the head groups.
Small permittivities facilitate the clustering of head groups with
a well defined maximum in the proton-proton pair distribution function $g(r)$ that indicates three simple multiplet
structures: compact and linear multiplets each created by two head
groups, and a branched cluster composed of three head groups. At higher
permittivity no correlations between head groups are obtained.
An effective repulsion between acidic head groups at intermediate
dielectric permittivities was revealed from our data. We assume that,
in the absence of strong correlations, each head-group dipole prefers
to stay in the bulk of the sample in order to increase its polarization
energy. 

For the ionomer interaction parameters considered in this work the
microphase structure of the head groups was found to have little dependence
on the concentration of backbone segments. Since the molar fraction
of sulfonates in a fuel cell membrane is known to be a crucial parameter
for its performance, this factor was also studied. We find that for
a dilute case a strong condensation of nearly all side chains takes
place. The main contribution to the Hamiltonian in this case comes
from a long-range dipole-dipole interaction of head groups which forces
the chains to aggregate. However, for a dense system, where the hydrophobic
LJ contribution to the Hamiltonian becomes comparable with the dipolar
term, the side chains cannot reorient themselves in order to minimize
the electrostatic energy of the aggregate. While on the one hand,
for high proton conductance one needs higher sulfonate molar fractions
$\eta$, on the other hand we find that a high value of this parameter
$\eta$ hinders the sulfonate aggregation that is vital for water-sulfonate
channel formation in ionomers.

Overall, our simulations show that the dipole-dipole interaction
is the main driving force for the aggregation of sulfonates. When
the dipole moment of sulfonates is small, so that  $D_r \ll 1$, there is no phase separation
in the ionomers. In contrast, for $D_r=1$ a microscopic aggregation
of hydrophilic and hydrophobic monomers takes place in the system.

The most significant feature learned from the simulations is the fact
that the dipolar-head-group approximation is not sufficient to drive
the compact side chain cluster formation suggested by experimental
measurements. An additional artificial attraction between the sulfonates,
introduced as a screened Yukawa interaction potential, boosts
formation of nanometer-sized clusters dispersed throughout the system
volume. The final size of these clusters, and the geometry of the
cluster network they create, depend on the dipole moment $D_r$ of the
sulfonate head groups for fixed Yukawa parameters. As the strength
of $D_r$ increases from zero to one, the conformational structure of
clusters changes from a Gierke model of spherical aggregations to
a Gebel model of elongated aggregations.

More work is required to develop a deeper understanding of the mechanisms
behind the aggregation picture observed experimentally. Because the
shape and size distribution of the aqueous pores in wet membranes
is not completely known, it will be useful to extend our preliminary
results for the dry Nafion considered in this paper to systems with
water present and with a more detailed chain structure. In future
work, we plan to extend the model presented here to more complex models.
In particular we will employ partial charges and Lennard-Jones parameters
\cite{jang2004,vishnyakov2001,wescott2006} for side chain and backbone
monomers of Nafion-like oligomers. The issues of how the partial charges
of ionomer molecules, in fixed and free proton models for side chains,
affect the nanophase structure of the PEM materials will be addressed.
A complete understanding of the role of head group charges on the
cluster conformations of side chains can provide helpful guidelines
for understanding existing membranes and designing new membrane materials
for fuel cells with improved characteristics.

\begin{acknowledgments}
This work was supported by the US Department of Energy under grant
DE-FG02-05ER46244, and was made possible by use of facilities at the
Case ITS High Performance Computing Cluster and the Ohio Supercomputing
Center.
\end{acknowledgments}

\newpage



\end{document}